\documentclass[draft]{agujournal2019}
\usepackage{url} %this package should fix any errors with URLs in refs.
\usepackage{lineno}
\usepackage[inline]{trackchanges} %for better track changes. finalnew option will compile document with changes incorporated.
\usepackage{soul}
\graphicspath{{./Figures}}
\draftfalse

\journalname{JGR: Space Physics}

\begin{document}

%% ------------------------------------------------------------------------ %%
%  Title
%
% (A title should be specific, informative, and brief. Use
% abbreviations only if they are defined in the abstract. Titles that
% start with general keywords then specific terms are optimized in
% searches)
%
%% ------------------------------------------------------------------------ %%

% Example: \title{This is a test title}

\title{Prediction and Verification of Parker Solar Probe Solar Wind Sources at 13.3~R$_\odot$}

%% ------------------------------------------------------------------------ %%
%
%  AUTHORS AND AFFILIATIONS
%
%% ------------------------------------------------------------------------ %%

% Authors are individuals who have significantly contributed to the
% research and preparation of the article. Group authors are allowed, if
% each author in the group is separately identified in an appendix.)

% List authors by first name or initial followed by last name and
% separated by commas. Use \affil{} to number affiliations, and
% \thanks{} for author notes.
% Additional author notes should be indicated with \thanks{} (for
% example, for current addresses).

% Example: \authors{A. B. Author\affil{1}\thanks{Current address, Antartica}, B. C. Author\affil{2,3}, and D. E.
% Author\affil{3,4}\thanks{Also funded by Monsanto.}}

\authors{S. T. Badman\affil{1}, % 0000-0002-6145-436X
         P. Riley\affil{2}, % 0000-0002-1859-456X
         S. I. Jones\affil{3,4}, % 0000-0001-9498-460X
         T. K. Kim\affil{5}, % 0000-0003-0764-9569
         R. C. Allen\affil{6}, % 0000-0003-2079-5683
         C. N. Arge\affil{4}, % 0000-0001-9326-3448
         S. D. Bale\affil{7,8}, % 0000-0002-1989-3596
         C. J. Henney\affil{9}, % 0000-0002-6038-6369
         J. C. Kasper\affil{10}, % 0000-0002-7077-930X
         P. Mostafavi\affil{6}, % 0000-0002-3808-3580
         N. V. Pogorelov\affil{5}, % 0000-0002-6409-2392
         N. E. Raouafi\affil{6}, % 0000-0003-2409-3742
         M. L. Stevens\affil{2}, % 0000-0002-7728-0085
         J. L. Verniero\affil{3} % 0000-0003-1138-652X
}

% \affiliation{1}{First Affiliation}
% \affiliation{2}{Second Affiliation}
% \affiliation{3}{Third Affiliation}
% \affiliation{4}{Fourth Affiliation}

\affiliation{1}{Center for Astrophysics $|$ Harvard \& Smithsonian, Cambridge, MA 02138, USA}
\affiliation{2}{Predictive Science Inc., San Diego, CA 92121, USA}
\affiliation{3}{NASA Goddard Space Flight Center, Greenbelt, MD 20771, USA}
\affiliation{4}{Catholic University of America, Washington, DC 20064, USA}
\affiliation{5}{University of Alabama, Huntsville, AL 35805, USA}
\affiliation{6}{The Johns Hopkins Applied Physics Lab, Laurel, MD 20723, USA}
\affiliation{7}{Physics Department, University of California, Berkeley, CA 94720, USA}
\affiliation{8}{Space Sciences Laboratory, University of California, Berkeley, CA 94720, USA}
\affiliation{9}{Air Force Research Laboratory, Space Vehicles Directorate, Kirtland AFB, NM 87117, USA}
\affiliation{10}{BWX Technologies Inc., Washington DC 20001, USA}

%(repeat as many times as is necessary)

%% Corresponding Author:
% Corresponding author mailing address and e-mail address:

% (include name and email addresses of the corresponding author.  More
% than one corresponding author is allowed in this LaTeX file and for
% publication; but only one corresponding author is allowed in our
% editorial system.)

% Example: \correspondingauthor{First and Last Name}{email@address.edu}

\correspondingauthor{S. T. Badman}{samuel.badman@cfa.harvard.edu}

%% Keypoints, final entry on title page.

%  List up to three key points (at least one is required)
%  Key Points summarize the main points and conclusions of the article
%  Each must be 140 characters or fewer with no special characters or punctuation and must be complete sentences

% Example:
% \begin{keypoints}
% \item	List up to three key points (at least one is required)
% \item	Key Points summarize the main points and conclusions of the article
% \item	Each must be 140 characters or fewer with no special characters or punctuation and must be complete sentences
% \end{keypoints}

\begin{keypoints}
\item Parker Solar Probe's orbit rapidly samples huge swathes of coronal structure leading to novel constraints on solar wind source regions.
\item Footpoint predictions for PSP are most accurate under quiet solar conditions with source locations concentrated in equatorial coronal holes.
\item We make an empirical determination of the Alfv\`en surface and relate protrusions in it to underlying coronal magnetic topology.
\end{keypoints}

%% ------------------------------------------------------------------------ %%
%
%  ABSTRACT and PLAIN LANGUAGE SUMMARY
%
% A good Abstract will begin with a short description of the problem
% being addressed, briefly describe the new data or analyses, then
% briefly states the main conclusion(s) and how they are supported and
% uncertainties.

% The Plain Language Summary should be written for a broad audience,
% including journalists and the science-interested public, that will not have 
% a background in your field.
%
% A Plain Language Summary is required in GRL, JGR: Planets, JGR: Biogeosciences,
% JGR: Oceans, G-Cubed, Reviews of Geophysics, and JAMES.
% see http://sharingscience.agu.org/creating-plain-language-summary/)
%
%% ------------------------------------------------------------------------ %%

%% \begin{abstract} starts the second page

\begin{abstract}
Drawing connections between heliospheric spacecraft and solar wind sources is a vital step in understanding the evolution of the solar corona into the solar wind and contextualizing \textit{in situ} timeseries. Furthermore, making advanced predictions of this linkage for ongoing heliospheric missions, such as Parker Solar Probe (PSP), is necessary for achieving useful coordinated remote observations and maximizing scientific return.  The general procedure for estimating such connectivity is straightforward (i.e. magnetic field line tracing in a coronal model) but validating the resulting estimates difficult due to the lack of an independent ground truth and limited model constraints. In its most recent orbits, PSP has reached perihelia of 13.3$R_\odot$ and moreover travels extremely fast prograde relative to the solar surface, covering over 120 degrees longitude in three days. Here we present footpoint predictions and subsequent validation efforts for PSP Encounter 10, the first of the 13.3$R_\odot$ orbits, which occurred in November 2021. We show that the longitudinal dependence of \textit{in situ} plasma data from these novel orbits provides a powerful method of footpoint validation.  With reference to other encounters, we also illustrate that the conditions under which source mapping is most accurate for near-ecliptic spacecraft (such as PSP) occur when solar activity is low, but also requires that the heliospheric current sheet is strongly warped by mid-latitude or equatorial coronal holes. Lastly, we comment on the large-scale coronal structure implied by the Encounter 10 mapping, highlighting an empirical equatorial cut of the Alfv\`{e}n surface consisting of localized protrusions above unipolar magnetic separatrices.
\end{abstract}

\section*{Plain Language Summary}
Parker Solar Probe (PSP) is a NASA heliospheric mission which travels closer to the Sun than any previous human-made object, but also is the first to fly faster than the Sun rotates and so skims over its surface in a new way. To get the most out of PSP's science and to tie its measurements of the solar wind back to the physics happening at the Sun, we need to estimate the solar origin of the plasma which later arrives at PSP. This paper describes how we predict these locations in advance of seeing the PSP data as well as how we check how well we did after the fact. We show how PSP's extremely fast motion across the Sun in recent orbits leads to new and powerful ways to verify our estimates. We show the conditions which make our predictions the most accurate is when there is low solar activity, but plenty of ``equatorial coronal holes'': dark regions seen near the Sun's equator where the Sun's plasma is able to escape into space. Lastly, we show how the connectivity exercise combined with the novel PSP orbital motion allows us to ``measure" a large portion of the Sun's atmosphere and relate these measurements to how the magnetic field lines of the Sun are arranged closer to the solar surface.

%% ------------------------------------------------------------------------ %%
%
%  TEXT
%
%% ------------------------------------------------------------------------ %%

%%% Suggested section heads:
% \section{Introduction}
%
% The main text should start with an introduction. Except for short
% manuscripts (such as comments and replies), the text should be divided
% into sections, each with its own heading.

% Headings should be sentence fragments and do not begin with a
% lowercase letter or number. Examples of good headings are:

% \section{Materials and Methods}
% Here is text on Materials and Methods.
%
% \subsection{A descriptive heading about methods}
% More about Methods.
%
% \section{Data} (Or section title might be a descriptive heading about data)
%
% \section{Results} (Or section title might be a descriptive heading about the
% results)
%
% \section{Conclusions}

\section{Introduction}\label{sec:introduction}
%Text here ===>>>
% Parker solar probe mission
Parker Solar Probe (PSP; \citeA{Fox2016}) is a NASA heliospheric mission launched in 2018 to explore closer to the Sun than ever before. PSP carries four instruments: The Electromagnetic Fields Investigation (FIELDS, \citeA{Bale2016}), the Integrated Science Investigation of the Sun (IS$\odot$IS, \citeA{McComas2016}), the Solar Wind Electrons, Alphas and Protons (SWEAP, \citeA{Kasper2016}) and the Wide-Field Imager for Solar Probe Plus (WISPR; \citeA{Vourlidas2016}). With the exception of WISPR, PSP carries only \textit{in situ} instrumentation. Further, WISPR is designed to view scattered white light well off the limb. As such, although PSP captures the local properties of the solar wind \textit{in situ} in exquisite detail, remote observations of where the plasma came from at its origin on the Sun must necessarily be made by other instruments and spacecraft on Earth and in the heliosphere or by ground-based telescopes. Making such connections back to the Sun is integral to all three of PSP's primary science objectives \cite{Fox2016}: 

\begin{enumerate}
    \item Trace the flow of energy that heats and accelerates the solar corona and solar
    \item Determine the structure and dynamics of the plasma and magnetic fields
at the sources of the solar wind
    \item Explore mechanisms that accelerate and transport energetic particles
\end{enumerate}

Many remote observatories, both space-based (e.g. the Interface Region Imaging Spectrograph, \citeA{DePontieu2014}) and ground-based (e.g. the Daniel K. Inouye Solar Telescope, \citeA{Rimmele2020}) have too small a field of view to observe the whole solar disk, or have high-resolution instrument modes with a smaller field of view (e.g. the Extreme Ultraviolet Imager \cite{Rochus2020} on board Solar Orbiter \cite{Muller2020}). As such, they require on-disk target coordinates as input to their nominal operations, and these coordinates must typically be provided hours to days in advance of the observation being made.

Therefore, to take full advantage of the array of solar observatories available, it is necessary to model the magnetic connectivity through the solar wind from PSP's location back to its origin low in the corona and derive target coordinates. Moreover, this modeling must be done as a \textit{prediction}, meaning it is done without any feedback in the form of \textit{in situ} data (section \ref{subsec:in-situ-data}) and using the most recent available boundary conditions which are assumed to be somewhat representative of the coronal conditions in the immediate future. Such predictions have been attempted by individual modelers since the mission launched \cite{Riley2019,Stansby2019b,vanderHolst2019,Kim2020,Wallace2022}.

The statement of the basic procedure for source mapping is deceptively simple. At its core, we consider the plasma between the source and the spacecraft (including the corona and the inner heliosphere or solar wind) to be well described at large scale by ideal magnetohydrodynamics (MHD) and that therefore the magnetic field is frozen in to plasma parcels as they flow away from the Sun. Thus, the time history and origin at the Sun of a given plasma parcel measured \textit{in situ} in the solar wind is simply estimated by magnetic field line tracing from the position of the measurement back towards the Sun (or identically, velocity streamline tracing in the solar-corotating frame, if the model includes the velocity field). See, e.g. \citeA{Stansby2022} for the governing equations for field line tracing of a vector field on a numerical grid. 

Despite the simple procedure, obtaining accurate results from this modeling task is far from trivial. Assessing its accuracy and uncertainty has been the goal of much recent work (e.g. \citeA{MacNeil2021,DaSilva2022,Koukras2022}). Our approach, which is the subject of section \ref{subsec:consensus}, is to gather model estimates from a diverse array of models (section \ref{subsubsec:models}), model parameters and physical boundary conditions, then form a distribution to assess the most probable source mapping and quantify the convergence. At several points in this work, the resulting mappings are illustrated using a potential field source surface (PFSS, \citeA{Altschuler1969,Schatten1969}) corona and a Parker spiral representation of the inner heliosphere \cite{Parker1958,Nolte1973, Neugebauer1998}. See, e.g., \citeA{Rouillard2020a,Badman2020,Koukras2022} and section \ref{subsubsec:models} for specific details of how the modeling works in this case. 

\begin{figure}
    \centering
    \includegraphics[width=0.9\textwidth]{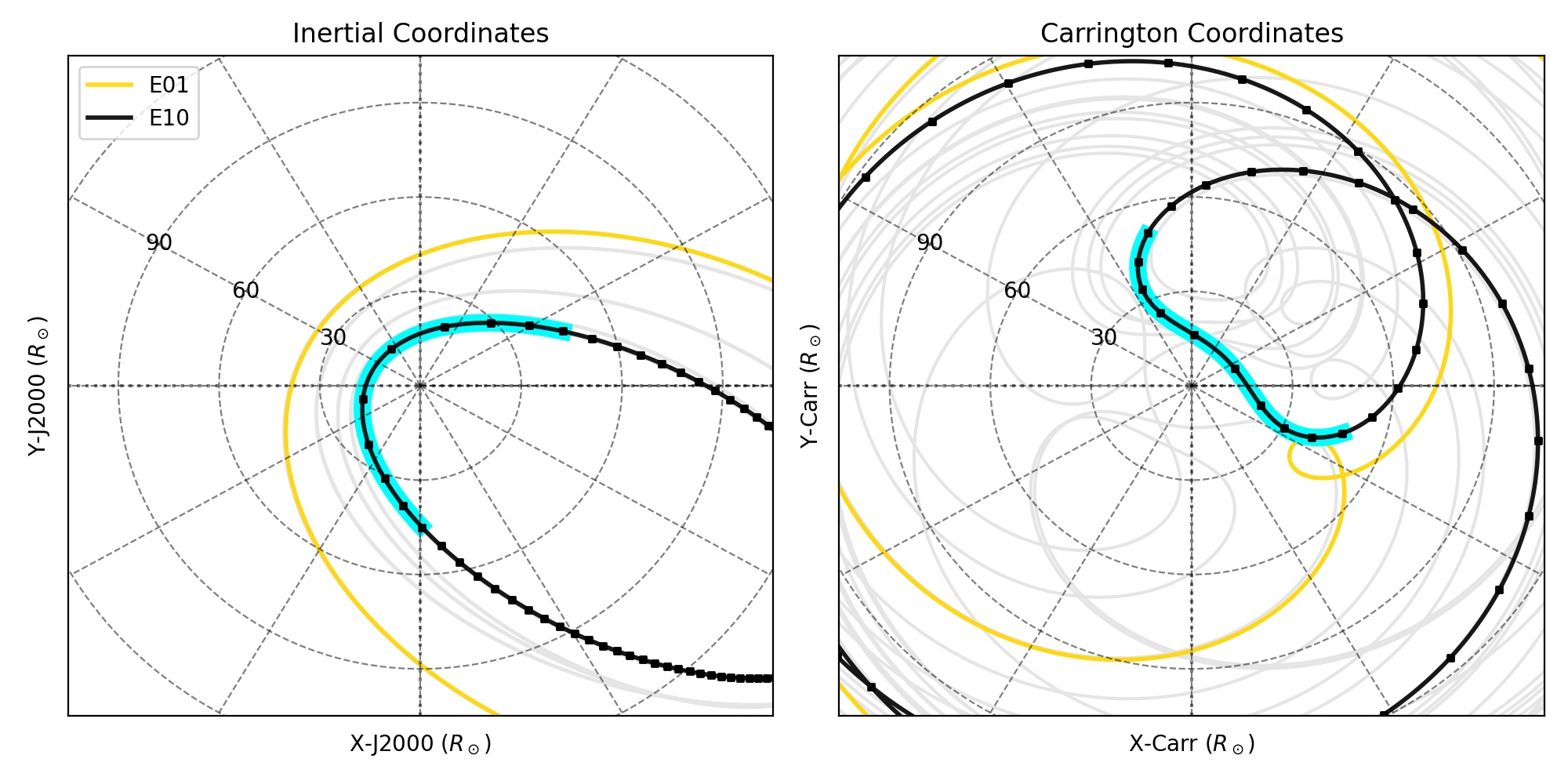}
    \caption{\textbf{Orbital Geometry of the 13.3$R_\odot$ orbit family.} PSP’s E10 orbital track projected into the solar equatorial plane is shown in black  with E01 shown in yellow for contrast. Black squares are spaced 24 hours apart. Faint grey curves show the full prime mission trajectory (i.e., 24 orbits).  The left-hand panel shows the inertial J2000 reference frame for which the orbit's elliptical nature and decreasing perihelion is clear. The right-hand panel shows the Carrington frame, which is most relevant for source mapping. This frame demonstrates the enormous range of solar longitude traversed by PSP in its latter orbits over just a few days around perihelion. Cyan shading shows the $\sim$9-day interval of E10 where PSP co-rotates or super-rotates with respect to the Sun, which is the focus of this study.}
    \label{fig:1}
\end{figure}

There are numerous novelties to performing these predictions for the PSP mission due to its unique orbits, whose key features we illustrate in figure \ref{fig:1}. In the left panel we show the mission trajectory in inertial coordinates for which the very elliptical nature of the orbits is clearly illustrated. In the right-hand panel, the same orbits are shown in the solar-corotating (or Carrington, \citeA{Carrington1863}) reference frame in which the Sun's angular rotation (25.38 day period) is removed, meaning only the spacecraft moves in these plots and solar wind stream structure can be plotted as static in time. This latter frame is, therefore, the most useful for understanding source mappings for PSP.

From these plots, we note that PSP’s orbits evolve significantly throughout the mission. They are organized into families of progressively closer perihelia separated by Venus gravity assists.  We refer to the portion of each orbit where PSP is close to the Sun and moves as faster or faster than solar rotation as an ``encounter'', and abbreviate e.g. ``Encounter 1'' as E01. 

In figure \ref{fig:1}, we contrast E01 (the very first perihelion of the mission from November 2018, which reached $35.7R_\odot$) and E10, the main subject of this manuscript which took place in November 2021 and was the first orbit to reach a perihelion of $13.3R_\odot$.
The contrast shows that even relative to the record-breaking first perihelion, PSP’s most recent orbits are dramatically closer to the Sun.

However, while the close approach distances are usually the headline of the mission, the angular motion in the co-rotating frame is equally as novel. The Carrington frame transformation reveals how PSP’s perihelia include an interval where PSP moves faster than solar rotation and therefore moves prograde with respect to sources on the Sun. In the most recent orbits, this period has expanded to cover over 150 degrees longitude, and the central 120 degrees longitude is covered in less than three days (see the black squares in figure \ref{fig:1}, which are spaced 24 hours apart). This leads to an unprecedented set of measurements covering a huge spatial range in a very short period of time. As we will see in this paper, this leads to a powerful new way to verify source mapping and infer large scale coronal structure directly from \textit{in situ} measurements.

The structure of this paper is as follows: In section \ref{sec:campaigns} we introduce the magnetic footpoint prediction campaigns carried out in support of each of PSP's perihelion passes. We summarize the methodology for collecting predictions and combining them to establish a consensus location. We then present in section \ref{sec:e10-results} an overview of the predictions made during PSP E10, the primary focus of this work. Next, in section \ref{sec:evaluation}, we turn our attention to utilizing the PSP data after each perihelion to evaluate the predictions and highlight how PSP's spatial motion in its most recent orbit makes this evaluation stronger. We then contrast the E10 results predictions with early predictions for E04 and show under which coronal conditions our predictions are most accurate and convergent (section \ref{subsec:e10-e04-comparison}). To close our analysis, we discuss how interpreting PSP measurements as large spatial cuts through coronal structure, we can use well-validated footpoint mapping to relate this structure to coronal magnetic topology and illustrate this with an example of a cut through the Alfv\`{e}n surface (section \ref{subsec:alfven-surface}).  In section \ref{sec:discussion}, we discuss the key inferences and takeaways from the various analyses presented in the prior sections.  Lastly, we close with sections \ref{sec:conclusions} and \ref{sec:future-work} where we present our major conclusions and identify directions for future work.  Additionally, in \ref{appendix:sec:practical}, as discuss miscellaneous practical aspects of making these predictions operationally and disseminating the results.

\section{PSP Footpoint Prediction Campaigns}\label{sec:campaigns}

Since its fourth solar encounter (E04, January 2020), PSP has spent a significant portion of its perihelia passages on the Earth-facing hemisphere of the inner heliosphere.  As a result, there has been a significant effort made to enable Earth-based observers to make coordinated observations of the specific locations on the Sun which source the plasma measured \textit{in-situ} by PSP (see \ref{appendix:sec:practical} for a discussion of the practical aspects of these campaigns). These locations will be referred to as ``(magnetic) footpoints'' or ``solar wind sources'' interchangeably throughout this work. Although such remote-\textit{in-situ} connections have been attempted for as long as there have simultaneously been both types of solar observatories online (e.g. \citeA{Nolte1976}), achieving it with PSP has the potential to be much more robust and informative due to its close approach to the Sun and the subsequent reduction in processing of the solar wind plasma between emission and receipt. Such targeted observations are of vital importance for maximizing the science return of missions like PSP, resulting in a two-way benefit in which remote observations can be used to contextualize the \textit{in situ} datasets, while the \textit{in situ} data can then be traced back to remotely observed processes on the Sun to better understand the intrinsic physical processes. As mentioned earlier, achieving localized observational targets ahead of time is important to use many of the highest resolution instruments at Earth since their fields of view are much smaller than the solar disk.

\subsection{Footpoint Predictions and Establishing a Consensus}\label{subsec:consensus}

A fundamental ingredient to achieving these coordinated observations is therefore making footpoint \textit{predictions} in advance of each orbit. Making these predictions necessarily entails computational modeling since the \textit{in-situ} data that can later be used for verification (section \ref{sec:evaluation}) has not yet been taken. For the PSP mission, this data is not even available near-real-time since the spacecraft attitude required to safeguard the system from solar radiation during perihelion precludes communication with Earth \cite{Kinnison2020}. The key challenge is how to produce the most accurate footpoints possible without \textit{in situ} data constraints.

The modeling involves producing global numerical representations of the coronal and heliospheric plasma, flying PSP's upcoming trajectory through the model, and tracing inwards continuous magnetic field lines (or equivalently under the frozen-in theorem, velocity streamlines in the co-rotating frame if the model includes fluid flow) that connect PSP to a certain location on the Sun which is specified as a heliographic coordinate on the solar surface (i.e., longitude and latitude at the photosphere). In this work, we utilize predictions from several different models with differing internal physics and input boundary conditions, but all are broadly carrying out this same high-level task (see section \ref{subsubsec:models} for a discussion of some of the differences between the models run for E10). 

\begin{figure}
    \centering
    \includegraphics[width=0.62\textwidth]{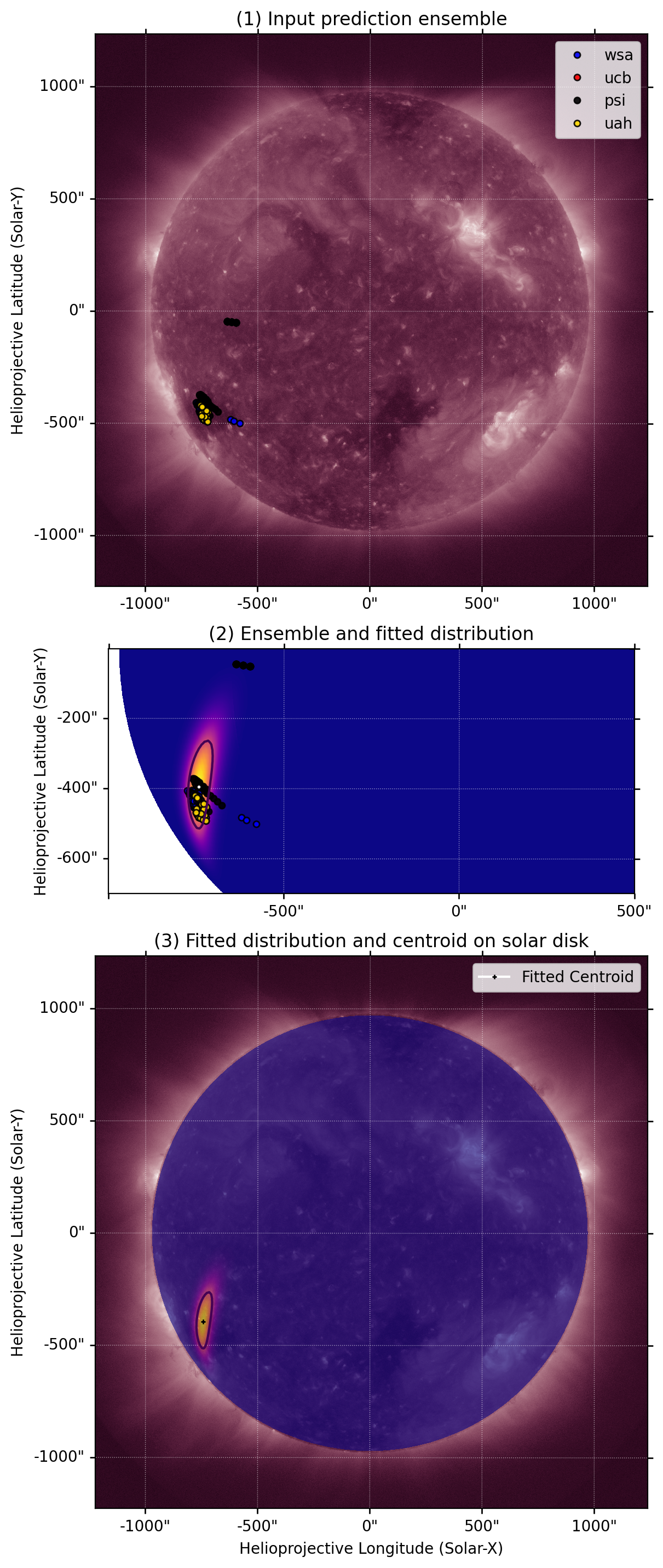}
    \caption{\textbf{Steps for establishing consensus from an ensemble of footpoint predictions.} (1) Footpoint predictions for a certain window of times of arrival at PSP supplied as Carrington longitude and latitude are gathered from different model predictions as an ensemble. (2) The ensemble is fitted to an elliptical gaussian on a sphere with a Kent distribution \cite{Kent1982,Fraenkel2014}. (3) From the fit, the distribution centroid is the ``consensus'' source location, while the half-maximum contour of the distribution encodes the uncertainty.}
    \label{fig:2}
\end{figure}

We use this universality of the modeling task as our primary method for maximizing prediction accuracy and quantifying the uncertainty: We simply require each model to produce a prediction in the same format: CSV files with columns ``Date'', ``Time'', ``Carrington Longitude'' and ``Carrington Latitude'' and data entries extending from the time of the prediction through the rest of the upcoming encounter (see \citeA{BadmanZenodo2023} for examples). We then collect the data from these files and bin the heliographic locations into windows of time of arrival of the plasma at PSP (initially every 6 hours in the first iterations of this methodology, but more recently every hour as PSP's angular velocity at perihelion has increased). We treat the collection of locations in each window as an ensemble of predictions from which we want to establish a consensus source location and record some information about the spread in predictions to encode the uncertainty.

This consensus establishment process is illustrated in figure \ref{fig:2}. Following from panels (1)-(3), we show how a given ensemble is drawn from several different models (and within each model, variation in model parameters and input boundary condition, i.e. there is a sub-ensemble for each model). The overall ensemble is then fitted with a Kent distribution (\citeA{Kent1982}, a.k.a. the 5-parameter Fisher-Bingham distribution or FB$_5$) , which is the spherical analogue of an elliptical gaussian in a 2D plane. This distribution behaves well on the whole sphere (i.e. it is not tripped up at the poles or by crossing periodic boundary conditions) and captures how concentrated the distribution is as well as the orientation on the sphere over which it is spread. The fit is performed with softwareimplemented by \citeA{Fraenkel2014}. Having fit the distribution, we then reduce the information down to the distribution peak, which we term the ``source centroid'', and a half-maximum contour (analogous to the full width at half-maximum of a 1D distribution) which is an ellipse encoding the uncertainty. The source centroid is interpreted as the ``consensus'' solar wind source that remote observers are advised to target (see section \ref{appendix:subsec:dissemination} for a discussion on the subtlety of the timestamp the target is valid for, a determination which has been actively evolving in recent iterations of this methodology).

\subsubsection{Models used for footpoint predictions for E10}\label{subsubsec:models}

The legend of figure \ref{fig:2} indicates the four different model sources for which predictions were provided during encounter 10. We briefly outline the differences and specifics of each model uses to derive footpoint estimates.

\begin{itemize}
\item \verb+wsa+ results come from the Wang-Sheeley-Arge (WSA, \citeA{Arge2000,Arge2003,Arge2004}) model. WSA combines the potential field source surface (PFSS, \citeA{Altschuler1969,Schatten1969}) and Schatten current sheet (SCS, \citeA{Schatten1972}) models to specify the global inner and outer corona magnetic field, respectively, and an empirical relationship to prescribe the asymptotic solar wind speed on a grid on the outer boundary of the coronal model, which is typically set to between 5 and 21.5Rs.  Solar wind macro-particles are then propagated quasi-ballistically (i.e., ad hoc stream interactions are included to prevent fast streams from bypassing slow ones) outward from the model boundary to the satellite, carrying with them the memory of the photospheric footpoint of the magnetic field line on which they originated.  For the PSP footpoint predictions, WSA models were run for ADAPT-GONG, ADAPT-HMI, and GONGz input magnetograms.
\item \verb+ucb+ results come from PFSS modeling performed with \verb+pfsspy+ \cite{Stansby2020} and ballistic mapping for the heliospheric part in a framework developed at UC Berkeley for supporting PSP's first orbit \cite{Bale2019,Badman2020}. In short, for a given magnetogram a PFSS model is run with some source surface radius $R_{ss}$. PSP's trajectory is then used to seed numerous Parker spiral field lines which are used to project PSP's position down to $R_{ss}$. At this point, field line tracing is performed using the built-in field line tracing tool in  \verb+pfsspy+ to draw a field line from $R_{ss}$ to $1R_\odot$. The longitude and latitude of this footpoint is then the prediction supplied. This process is iterated with different global magnetic maps (ADAPT-GONG and ADAPT-HMI, and ADAPT realizations therein \cite{Arge2010,Arge2011,Hickmann2015}) and values of $R_{ss}$ for a given prediction day.
\item \verb+psi+ predictions come from models run at Predictive Science Inc., which include (1) an independent PFSS solver \cite{caplan2021} and ballistic heliosphere, with steps similar to those of \verb+ucb+, and (2) predictions from the Magnetohydrodynamics Algorithm outside a Sphere (MAS) (e.g., \citeA{Riley2019,Riley2021e}). For the MHD solutions, we map the solar wind ballistically back to $30 R_S$, after which we trace along the appropriate field lines back to the solar surface. When the spacecraft is within $30 R_S$, there is, of course, no ballistic component to the mapping procedure. As with \verb+ucb+, the PSI team creates a suite of realisations using both ADAPT-GONG and ADAPT-HMI magnetograms (updated to 1200 and 0800 on the day of the prediction, respectively), together with a range of source surface heights for the PFSS model. 
\item \verb+uah+ predictions come from the University of Alabama, Huntsville Multiscale Fluid-Kinetic Simulation Suite (MS-FLUKSS, \citeA{Pogorelov_XSEDE,pogorelov2023iau,Singh_2022}), which can solve the Reynolds-averaged ideal MHD equations for the mixture of thermal and nonthermal solar wind ions coupled with the kinetic Boltzmann equation describing the transport of neutral atoms. An adaptive mesh refinement technique can be employed for efficient high-resolution calculations. The MS-FLUKSS heliospheric MHD model is coupled with the WSA model \cite{Kim2020}, which uses both ADAPT-GONG and ADAPT-HMI input magnetograms, with the PFSS source surface height and the WSA outer boundary at $2.5_\odot$ and $10_\odot$, respectively. Hence, field line tracing is performed through the MHD domain down to $10 R_\odot$ instantaneously at approximately 1 hour cadence, where the origin of the field line on the photosphere is already known, as described earlier for WSA.
\end{itemize}

% here's the reference for caplan2021...can't see it in the comments
%@article{Caplan2021,
% doi = {10.3847/1538-4357/abfd2f},
% url = {https://doi.org/10.3847/1538-4357/abfd2f},
% month = jul,
%  publisher = {American Astronomical Society},
%  volume = {915},
%  number = {1},
%  pages = {44},
%  author = {Ronald M. Caplan and Cooper Downs and Jon A. Linker and Zoran Mikic},
%  title = {Variations in Finite-difference Potential Fields},
%  journal = {The Astrophysical Journal}
%}

\section{E10 Footpoint Predictions}\label{sec:e10-results}
In the present work, we are primarily focused on the results of the prediction campaign for Encounter 10 (E10) which took place in November 2021, and was the first PSP orbit to reach a perihelion distance of 13.3$R_\odot$, (the prior distance was $15.8R_\odot$ as of April 2021). As discussed in section \ref{sec:introduction}, in addition to the substantially closer perihelion distance, the newest orbits of PSP are also novel in their extremely large longitudinal coverage around the Sun in a very short time over perihelion.

\begin{figure}
    \centering
    \includegraphics[width=\textwidth]{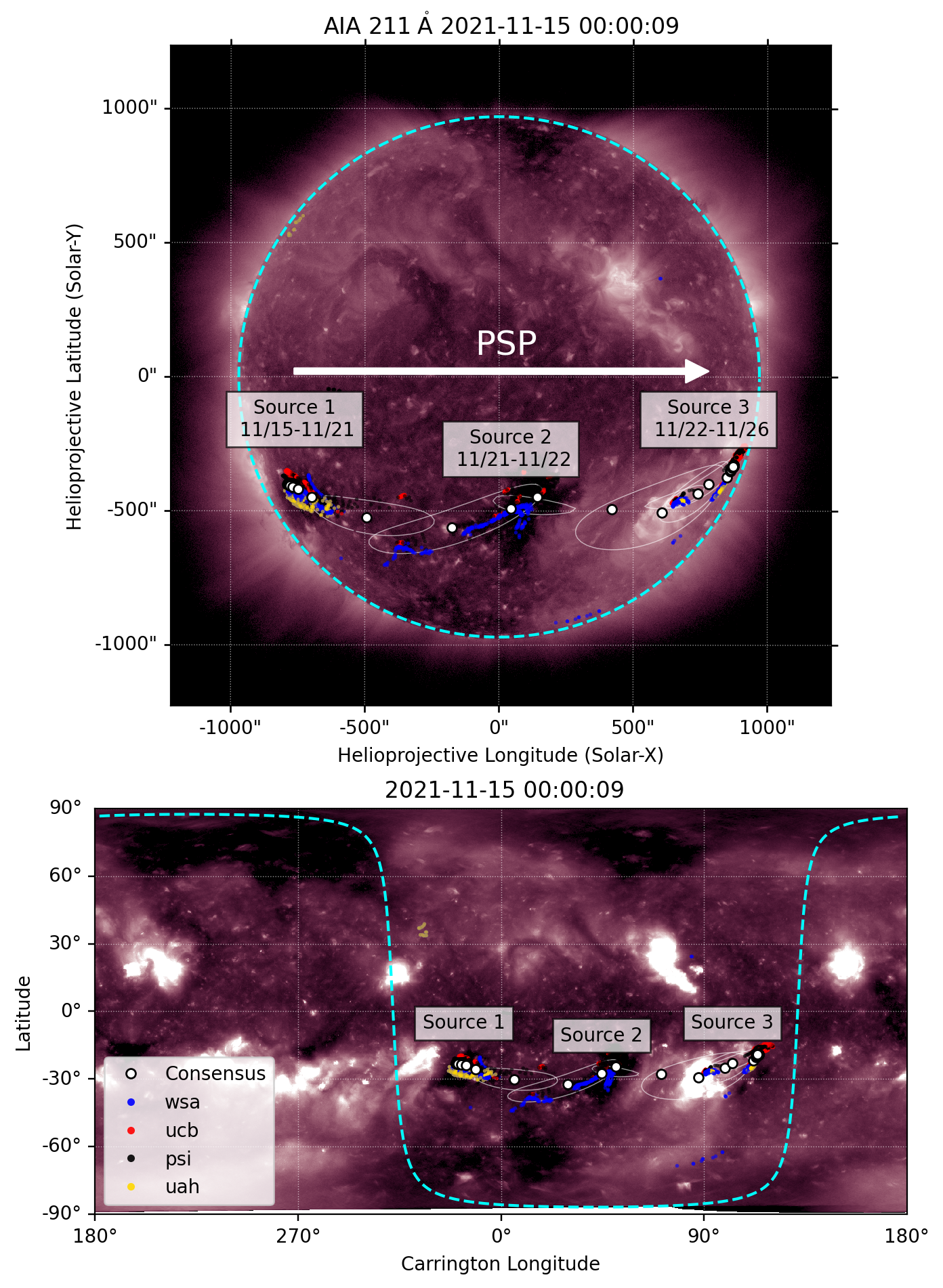}
    \caption{\textbf{Summary of footpoint predictions for PSP Encounter 10}.  Both panels depict the encounter 10 predictions for PSP’s footpoint connectivity as updated on 2021/11/15, superimposed on SDO/AIA 211 \r{A} remote observations. The top panel is the helioprojective view from the date of the update, while the bottom panel shows a heliographic projective with EUV data from a Carrington rotation centered around the date of the update. Different colored dots show the distribution of predictions from different models (see main text), while the white points with black outlines show the consensus, and the faint white contours show the associated half-maximum contour (see section \ref{subsec:consensus}). The footpoints track from left to right in time in both plots as PSP rotates faster than the Sun during perihelion. 
}
    \label{fig:3}
\end{figure}

E10 was a ``partially on-disk'' encounter where PSP's footpoints passed behind the west limb shortly after perihelion, and therefore had a small-scale prediction campaign (section \ref{appendix:subsec:scope}). Updates were issued on 2021/11/8, 2021/11/11 and 2021/11/15. Perihelion occurred on 2021/11/21 at 8am UTC, while limb passage was predicted to occur by 12pm UTC the same day. 

In figure \ref{fig:3} we present the source predictions and derived consensus source locations produced at the time of E10, i.e. prior to \textit{in situ} data from that orbit becoming available for the last issued update for the encounter (2021/11/15). The two-panel figure shows the input ensembles as colored scatter points (refer to section \ref{subsubsec:models} for specifics on the ensemble members). The derived consensus and uncertainty ellipses are shown as white markers and white contours projected in both the helioprojective frame as seen from Earth on 2021/11/15 (top) and in the Carrington frame (bottom). The background texture is SDO/AIA 211 \r{A} Extreme ultraviolet data (For the Carrington frame it is assembled from 1 solar rotation centered on 2021/11/15). A cyan dashed curve shows the same solar limb contour in both panels allowing the panels to be related to each other.

The top line of the prediction is straightforward to see from the figure. As annotated, the predicted footpoints (which move from left to right chronologically in both panels as PSP rotates faster than the solar surface) cluster in turn in a series of three isolated mid-latitude coronal holes in the southern hemisphere. All three are verifiable as dark regions in the EUV data. The first two have significant spatial extent, while the third is thin and extended. As will be verified later with magnetic modeling, all three are negative polarity and therefore the prediction also entailed suggesting PSP would not have any heliospheric current sheet crossings during this encounter.

Figure \ref{fig:3} also emphasizes the large range of solar longitude crossed by PSP during its latest encounters over a relatively short time. In fact, the entire traversal from one limb to the other shown in this plot takes place over just 9 days, with the bulk of that time occupied by PSP hovering over source 1 and source 3 as it respectively goes through inbound and outbound corotation with the Sun (see the date labels in the figure). We note that a subtlety not directly conveyed by the figure that although all three sources are visible on disk on the date the prediction was issued, the Sun continues to rotate as seen from Earth. As a result, source 3 rotates behind the limb before PSP reaches it. However, the first two sources were synchronously visible while PSP connected to them.

The predictions are also tightly clustered. For time intervals where PSP is predicted to connect to one of the three coronal holes, almost all ensemble members were located within the same coronal hole. There was no bifurcation. This led to high confidence in this prediction despite the lack of available \textit{in situ} data at the time it was issued.

Note that of the consensus points (white circles) \textit{not} located within the coronal holes, their location is driven by difficulties in the consensus fitting procedure when PSP is moving extremely rapidly across the disk. For this encounter, we still used six-hour windows for the consensus fitting. However, for this orbit, six hours near perihelion constitutes a large amount of longitudinal motion which causes some windows to contain ensemble members from two of the sources. When generating the consensus then, the fitted centroid was averaged between the sources, and the error ellipse grew large and protracted in longitude, as can be verified from the figure.

\section{Prediction Evaluation}\label{sec:evaluation}

 We now combine these predictions with \textit{in-situ} data obtained after the encounter. We discuss how we cast the data to spatial coordinates and then use it to evaluate the predictions, thereby inferring the global coronal structure. 

\subsection{In-situ Data and Ballistic Projection}\label{subsec:in-situ-data}

\begin{figure}
    \centering
    \includegraphics[width=\textwidth]{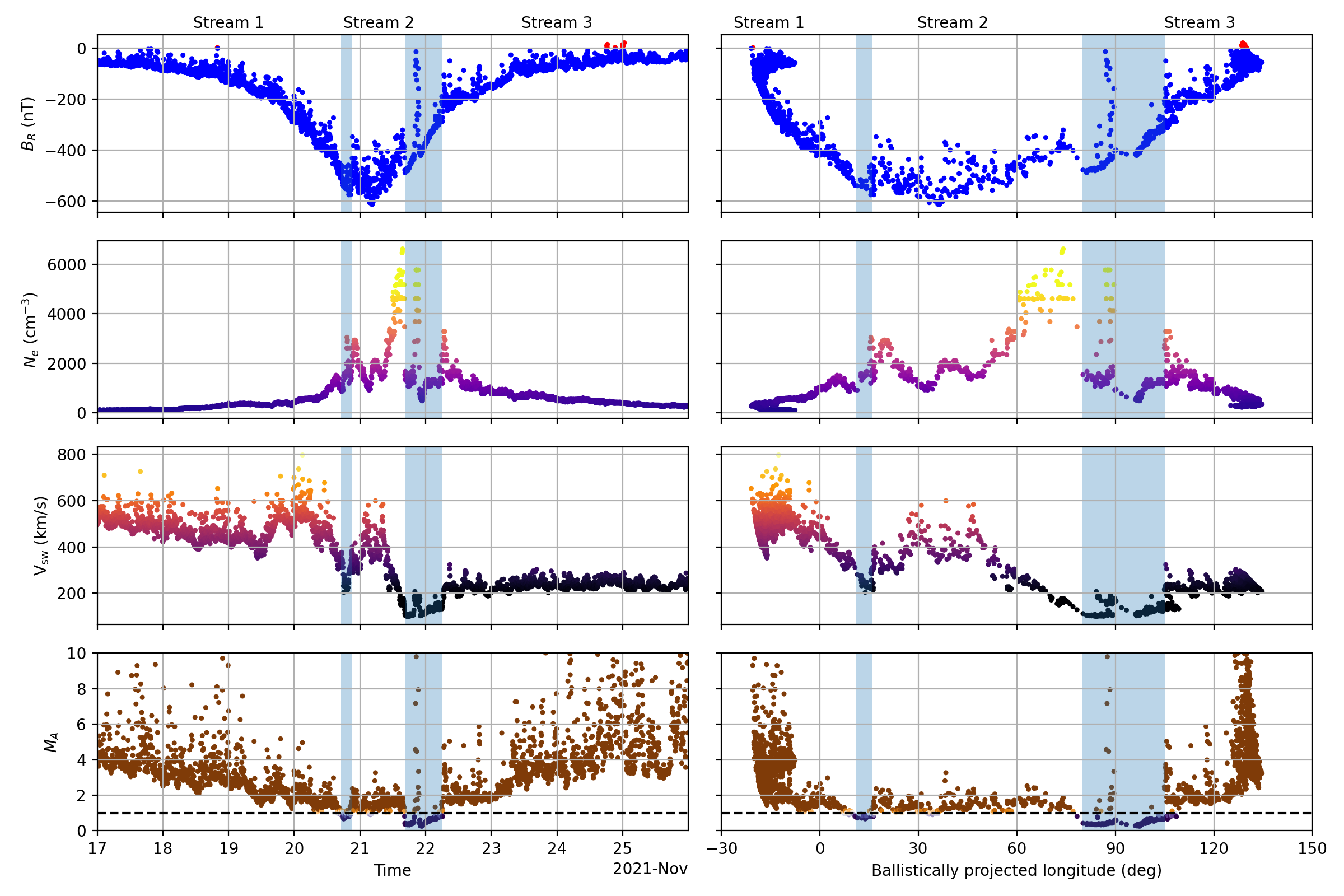}
    \caption{\textbf{Temporal and spatial distribution of selected \textit{in-situ} data taken during PSP Encounter 10.}  From top to bottom, data shown are the radial component of the magnetic field vector ($B_R$, PSP/FIELDS), QTN-derived electron density ($N_e$, PSP/FIELDS/RFS), the radial component of the solar wind velocity ($V_\mathrm{SW}$, PSP/SWEAP/SPAN-ion, proton channel moment), and the Alfv\`en mach number ($M_A = V_\mathrm{SW}/V_A$) where $V_A = B_R/\sqrt{\mu_0 m_p N_e}$ is the Alfv\`en velocity). The left-hand panel shows the measured timeseries interpolated to a constant cadence of 5 minutes while the right-hand panel shows the same dataset but as a function of spacecraft position ballistically projected down to 2.5$R_\odot$, as described in the main text. Colormaps for the bottom three rows highlight the relative variation of the relevant data set and allow features to be identified in both the left and right-hand columns. The colormap for the $B_R$ profile (blue-red) simply encodes the magnetic polarity, which is essentially unipolar and negative.
}
    \label{fig:4}
\end{figure}

We first introduce \textit{in-situ} data taken by PSP during encounter 10, specifically looking at the nine-day interval over which PSP was corotating or super-rotating with respect to the corona as highlighted earlier in figure \ref{fig:1}. In the left-hand column of figure \ref{fig:4} we show (from top to bottom) timeseries  of the radial component of the magnetic field ($B_R$) measured by PSP/FIELDS \cite{Bale2016}, the electron density ($N_e$) as measured by quasi-thermal noise (QTN) signatures \cite{Moncuquet2020} from PSP/FIELDS/RFS spectrograms \cite{Pulupa2017} and the solar wind radial velocity moment ($V_R$) from PSP/SWEAP/SPAN-i \cite{Kasper2016,Livi2022}. In the bottom panel, we compute the Alfv\`en mach number $M_A = V_R/V_A$ where $V_A=B_R/\sqrt{\mu_0 m_p N_e}$ is the Alfv\`en speed. The bottom panel is therefore a derived quantity from the 3 panels above. This derivation requires that the input data are all evaluated at the same timestamps. We therefore linearly interpolate all this data to a five-minute cadence, which is plotted in figure \ref{fig:4}.

Timeseries are the most common format in which \textit{in-situ} heliospheric datasets are typically presented, and for PSP they can still be used to extract immediate interesting properties of the encounter. For example, the radial magnetic field shows PSP remained in a negative polarity stream throughout the encounter, and its magnitude scaled with the usual $1/R^2$ Parker spiral trend \cite{Bale2019,Badman2021}. The minimum in $B_R$ represents the time of perihelion in the timeseries. The electron density is also seen to grow around perihelion. The velocity meanwhile shows that PSP measured moderately fast solar wind speeds in the inbound phase and then measured slow wind for much of the outbound period. The Alfv\'en mach number shows a smooth large-scale variation which is driven by the steady increase in the Alfv\`en speed as the magnetic field strength increases. Near to perihelion, it is close enough to $M_A=1$ that variation in stream structure drives excursions below the Alfv\'en surface. However, we remark that judging just from the timeseries, the significance of these periods may be underestimated as explained in the next paragraph. We will explore this Alfv\`en surface structure further in section \ref{subsec:alfven-surface}.

However, to fully benefit from PSP's unique measurements, it is more powerful to cast the timeseries data returned by the spacecraft into spatial coordinates which co-rotate with the Sun. To achieve this, and subsequently to compare to coronal model footpoint predictions, we produce the PSP trajectory for a set of timestamps over the time period of interest (the same set used to interpolate the data for computing $M_A$). We then use the measured velocity at each interpolated timestamp to ballistically project \cite{Nolte1973,Badman2020,MacNeil2021} PSP's position at that time to a specific Carrington longitude (and latitude) at 2.5$R_\odot$. PSP's unique orbit means this is not a linear or intuitive transformation. It causes the parts of the timeseries around corotation to collapse or fold back on themselves, while it stretches out periods near perihelion when PSP's angular motion is maximal, and thus reveals that structures or streams which might appear insignificant in the time series in fact have a substantial spatial extent.

Having applied the projection, we can now plot our interpolated datasets with the mapped longitude as the x-axis and observe how the measurements are distributed across the Sun. The results are shown in the right-hand column of figure \ref{fig:4}. As alluded to in the previous paragraph, the transformation strongly deforms the timeseries. We see that the small time period near perihelion, where field magnitude and Alfv\`en mach number are relatively flat, constitutes most of the spatial motion during the encounter. The two sub-alfv\`enic intervals which appeared small in the time series are actually respectively 5 and 20 degrees in longitudinal extent. 

The most interesting transformation for the present work, however, is to the velocity profile. We see that it collapses into three distinct streams of similar longitudinal extent. The first two streams are relatively fast (400-500 km/s) while the last one is very slow ($\sim$200 km/s). Traversing between each of the streams the solar wind speed drops and in both cases becomes sub-alfv\`enic for some or all of the transition. Recall that the general prediction for this encounter (without seeing the data) was for three sequential negative polarity coronal hole sources, the first two with significant areal extent while the third was narrow and extended. 

\subsection{Data-Model Comparison}\label{subsec:e10-eval}

\begin{figure}
    \centering
    \includegraphics[width=\textwidth]{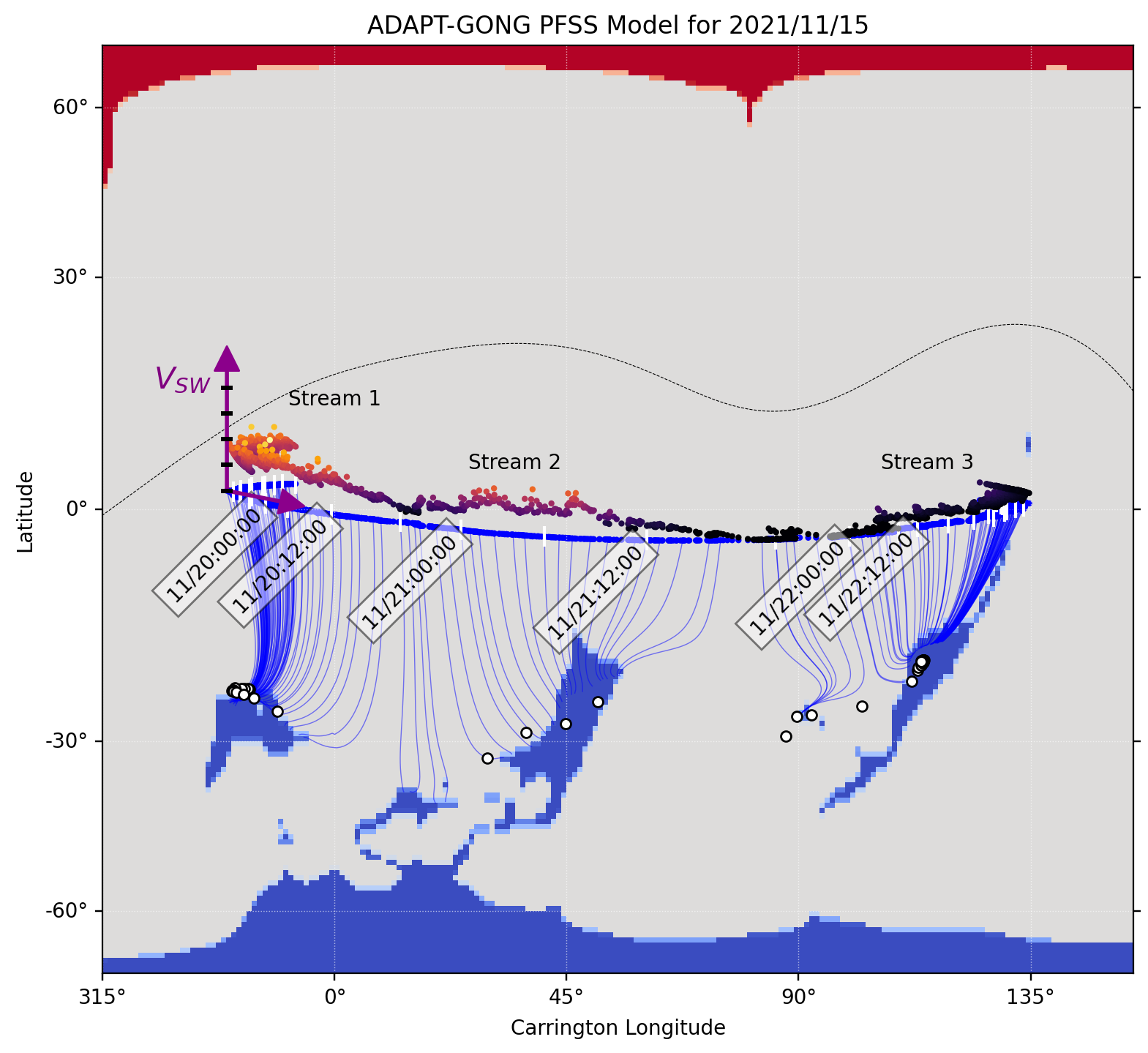}
    \caption{\textbf{Correspondence between in situ stream structure and magnetic field mapping.} A representative PFSS model (ADAPT-GONG 2021/11/15, 2.5Rs source surface height) is used to generate field lines connecting the ballistically mapped PSP trajectory (blue, indicating consistent measured negative polarity by PSP/FIELDS) to its sources. Blue/red shading indicates the locations of coronal holes in this model. A narrow dotted line indicates the neutral line/heliospheric current sheet from the model. White markers indicate the consensus footpoint locations shown in Fig 3. The dark/light colored trace depicts the PSP/SWEAP/SPAN-i data shown in Fig 4. as a function of source surface longitude, allowing the thermal plasma stream structure to be visually associated with the source mapping . It’s variation is plotted relative to the latitude of PSP’s trajectory. White vertical bars indicate six-hour separation in the trajectory as PSP moves from left to right. 
}
    \label{fig:5}
\end{figure}

In figure \ref{fig:5} we illustrate the correspondence between the data introduced above and the footpoint predictions with the use of a representative PFSS model (meaning a model where the footpoints are near to the consensus points) to supply field lines and a model heliospheric current sheet (HCS). The HCS is located well above the latitudes covered by PSP in its orbit, which explains why the predictions are unipolar and negative polarity, and is consistent with the measured polarity as indicated by the colorization (blue) of PSP's trajectory in the figure. 

However, E10 also yields a ready comparison between the solar wind velocity streams measured \textit{in situ} and the predicted sources. This is illustrated in Figure \ref{fig:5} by the colored scatter points superimposed above the blue perihelion loop. These are annotated such that zero velocity would lie on the blue trajectory curve, and $V_R>0$ causes a displacement upwards in latitude. This shows how the solar wind velocity spatial structure presented in figure \ref{fig:4} directly corresponds to the source mapping between the three distinct coronal holes. To put it explicitly, the peaks in solar wind velocity in each stream occur when the source mapping is in the center of each coronal hole, while it slopes off towards the edges.

Figure \ref{fig:5} also illustrates how around the corotation intervals, the timeseries fold back on themselves. The upshot of this is that the initial fast wind data in figure \ref{fig:4} from November 17-20 all correspond to the same coronal hole source: Its source first moves slowly eastwards over the center of the stream (but not reaching the east-most extent, since the velocity stays high), then reverses westwards. Beyond this point it continues on to the next two sources through the remainder of the encounter. A similar turnover is seen above source 3 for the outbound phase, but for continuously slow wind. Both intervals may therefore be regarded as ``fast radial scans'', where the inbound is in fast wind, and the outbound is in slow coronal hole wind. Such scans are very interesting targets for studies that aim to study the same stream at multiple heliocentric distances (e.g. \citeA{Mashayekhi2023}).

\subsection{Encounter 04 Comparison - Coronal Conditions for Accurate Source Predictions}\label{subsec:e10-e04-comparison}

\begin{figure}
    \centering
    \includegraphics[width=\textwidth]{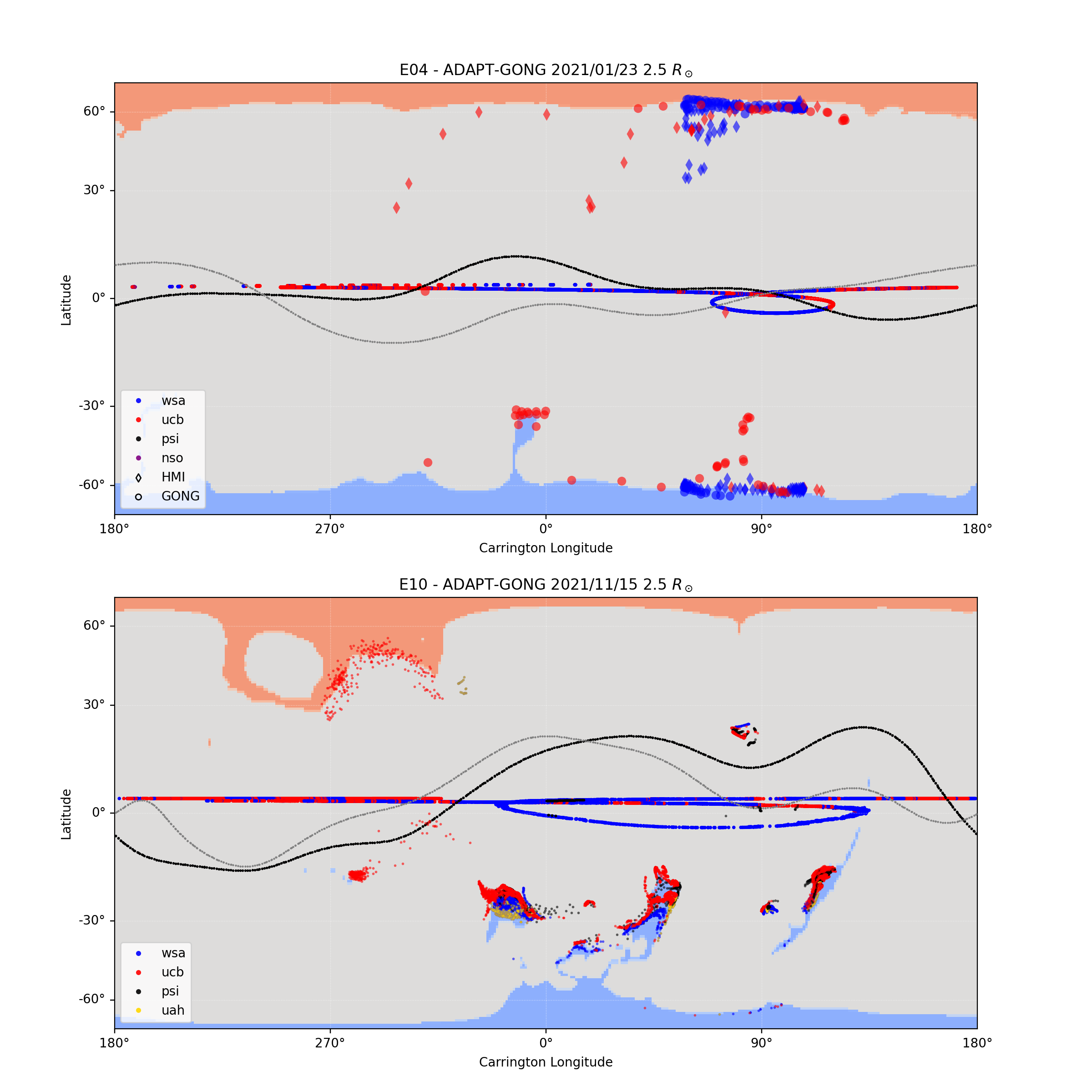}
    \caption{\textbf{Footpoint prediction distribution for different coronal configurations.} Top panel: The footpoint predictions made for PSP Encounter 04 (January 2020) from different models and input magnetograms. The extremely flat HCS in the vicinity of PSP’s perihelion leads to strongly bifurcated and unstable predictions split between both polar coronal holes, and bifurcation between models driven by ADAPT-GONG (circles) vs ADAPT-HMI (diamonds) -based magnetograms. Bottom panel: E10 footpoint predictions (see fig 3.). A fainter grey HCS contour in both panels show how the ADAPT-HMI neutral line for the same magnetogram date differs from the equivalent ADAPT-GONG one.
}
    \label{fig:6}
\end{figure}

In the previous section, we showed multiple independent lines of evidence that our prediction campaign for Encounter 10 was highly successful. In this section, we briefly contrast this encounter with another to comment on the coronal conditions under which footpoint predictions are most accurate. 

It is evident that time-dependent phenomenon will disrupt the accuracy of our footpoint estimates, since the models we run (see section \ref{subsubsec:models}) are all time-independent. Further, they explicitly assume that the current state of the corona will remain the true state for some days into the future, since this is necessary to run predictions. Such time-dependence is ubiquitous in times of high solar activity. As such, we might assume that the condition for accurate predictions should be intervals of low solar activity. However, here we show that ``low solar activity'' is not a sufficient condition for accurate estimates.

In figure \ref{fig:6} we present prediction ensembles (not consensus, but all modeler contributions) for Encounter 04 (January 2020, top panel) and Encounter 10 (November 2021, bottom panel) in the Carrington frame. We display the PSP perihelion trajectory for both encounters and color by the measured polarity (blue: $B_R < 0$, red : $B_R > 0$). We also choose a representative magnetogram date for both encounters and derive (via PFSS modeling) a HCS curve using both GONG and HMI-derived ADAPT global magnetic maps (respectively, black and grey curves in the plots) and a coronal hole map from the ADAPT-GONG magnetogram (modeled open field locations). In E04, we also keep track of GONG and HMI-based predictions by differentiating the markers as circles and diamonds respectively.

We observe that during Encounter 04, the predictions were extremely bifurcated: All models (different colors) were predicting portions of their ensembles to be in the northern or southern polar coronal holes at similar longitudes. Not only was this instantaneously true for the prediction depicted in figure \ref{fig:6}, but additionally the ensemble members from the same magnetogram type could switch hemisphere from one prediction update to the next as the most recent magnetograms were updated. Typically during this campaign, footpoints derived from GONG-based magnetograms clustered in one hemisphere and HMI-based magnetograms clustered in the other. This is most obviously seen in figure \ref{fig:6} around 0 longitude where diamonds (HMI) are all in the northern hemisphere, and circles (GONG) are in the southern hemisphere. Although harder to see, this also occurs in the perihelion loop (around 90 degrees) where the footpoints are denser. 

The root issue here was that the coronal conditions for this encounter were extremely quiet and solar minimum-like: The open magnetic field was all confined to the polar coronal holes and the resulting HCS curve was very flat and almost perfectly equatorial. Due to PSP also orbiting very near the heliographic equator, this meant that only a small shift in the model HCS line would move PSP's location from one side of the streamer belt to another. These shifts were small enough to be caused by the normal variation of changing the on-disk portion of the magnetogram by one day and, as illustrated by the black vs. grey HCS, by varying which instrument is used to assemble the magnetogram even if issued on the same day.

Additionally, even assuming predictions did converge on one hemisphere or another, when the only source is the edge of the featureless polar coronal hole, it is hard to be accurate since predictions essentially follow the longitude of the ballistically propagated field line. Thus there is a large spread in longitudes for a given prediction window which is constrained not by the well-modeled coronal hole topology, but rather by the crudely modeled heliospheric field lines.

The contrasting situation for E10 can be seen from the bottom panel of figure \ref{fig:6}. As we have reported in prior sections, the magnetic footpoints from all different modeling sources and input boundary conditions clustered very tightly on single well-defined sources. This occurs because the heliospheric current sheet is tilted up and away from the orbit of PSP significantly enough that small perturbations in the model from differing magnetograms (again, visually compare the black and grey curve) do not cause any magnetic separatrices to cross PSP's trajectory at perihelion. Further, because each source was an isolated mid-latitude coronal hole, each expands in the model to cover a large region in the outer corona, so that a large portion of PSP's orbital path maps down to a much smaller region. This means that instead of the spread in footpoints being typically set by the spread in ballistically-mapped longitudes, it is instead set by the size of the coronal hole itself, which is much smaller \cite{Koukras2022}. 

Finally, these contrasting campaigns also show how the much larger corotation loop in the most recent orbits also helps the situation by causing the connectivity to move between very different sources in the same perihelion. Thus the signatures of source changes, as shown in previous sections, are much more obvious in the \textit{in situ} data.

Thus we conclude that the most robust situation for source mapping (for a near-ecliptic spacecraft) is not the most quiet, solar minimum-like conditions, since a flat featureless corona gives rise to difficulties navigating the HCS separatrix when trying to predict footpoints. Instead, we do indeed benefit from low solar activity in that it negates the impact of time-dependent phenomena, but we also require that there be strong perturbations to the heliospheric current sheet, and correspondingly, mid or equatorial coronal hole sources which cause such perturbations. Both the presence of the warped current sheet and isolated coronal holes contribute to reduced uncertainty in our footpoint predictions.

\subsection{Inferring Global Coronal Structure : The Alfv\`en Surface of E10}\label{subsec:alfven-surface}

In this last analysis section, we demonstrate how combining the \textit{in situ} measurements with the verified field line mapping allows powerful inferences of global coronal structure.

\begin{figure}
    \centering
    \includegraphics[width=0.7\textwidth]{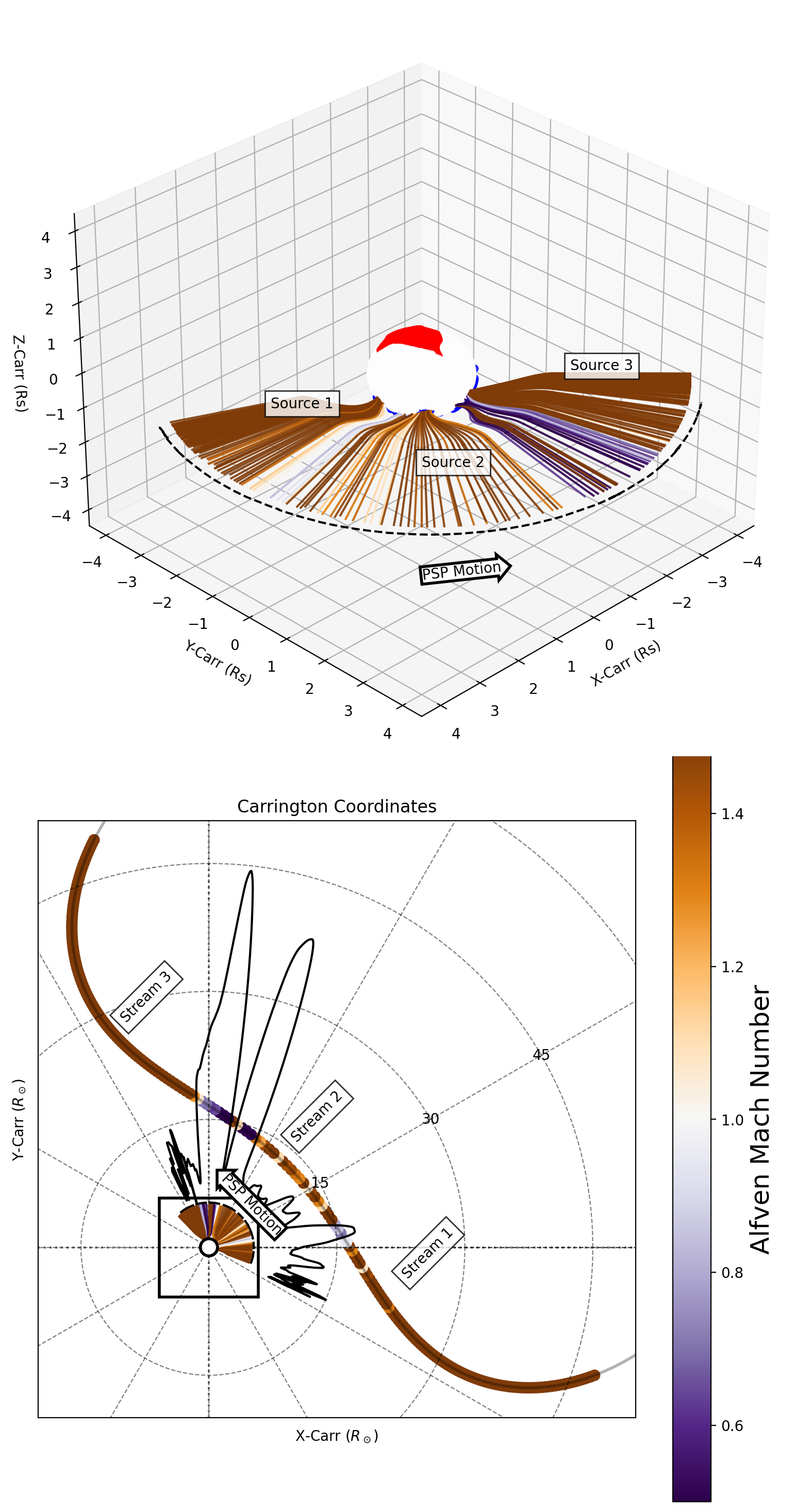}
    \caption{\textbf{Empirical constraints on the shape of the Alfv\`en surface and source region relations.} The bottom panel shows PSP’s E10 trajectory (bottom right to top left in time), colorized by measured Alfv\`en Mach number. An empirical estimate of the Alfv\`en surface constrained by the measured $M_A$=1 crossings is shown in black (see main text). A small square shows the region depicted in the top panel which shows PFSS field line mappings from 4~$R_\odot$ down, colorized by the measured Alfv\`en mach number so they can be related to the Alfv\`en surface structure above. Both instances of sub-alfv\`enic wind (corresponding to an outward protrusion in the Alfv\`en surface) are associated with pseudostreamer magnetic topology as in \citeA{Kasper2021}, although the first instance also appears to coincide with a traversal of a CME leg \cite{McComas2023}.
}
    \label{fig:7}
\end{figure}

In the bottom panel of figure \ref{fig:7} we display PSP's Carrington-frame trajectory for E10 projected into the heliographic equatorial plane and colorize it by the Alfv\`en mach number following the same color normalization as in figure \ref{fig:4}, which clearly shows the spatial extent and locations of the two sub-alfv\`enic intervals. 

On top of this, we superimpose a simple construction to create a visual representation of the topology of the Alfv\`en surface. Following \cite{Liu2021,Liu2023-preprint}, we take PSP's heliocentric distance as a timeseries and divide it by $M_A$ where we smooth the data via a rolling 20 minute (4 $\times$ 5 minute data points) window. Recombining this modified radius coordinate with PSP's Carrington longitude gives us the black curve which necessarily intersects PSP's orbit when $M_A=1$, is further out than PSP when $M_A < 1$ and closer in that PSP when $M_A > 1$.  The curve is not intended to be quantitative in how much further out or in it is relative to PSP when $M_A \neq 1$, but does qualitatively capture the large-scale undulations in the surface which are implied by the data.  However, we note that this construction as used by \citeA{Liu2021,Liu2023-preprint} is the limiting case of physical scaling assuming the radial magnetic field and plasma density vary as $1/r^2$ and a constant solar wind speed (e.g. \citeA{Weber1967}). In particular, the latter assumption of constant wind speed needs quantitative re-examination with PSP now measuring substantial solar wind acceleration \cite{Dakeyo2022,Halekas2022}, but for the purposes of the present manuscript is sufficient to describe when PSP is above or below the Alfv\`en surface.

Thus we see the PSP data combined with its unique orbit allow estimation of large-scale properties of the corona and imply that the Alfv\`en surface is typically within 13$R_S$ but has protrusions outwards which PSP has crossed. This empirical finding is interesting in its own right and consistent with a ``wrinkled surface'' as in \citeA{Liu2023-preprint}, as well as the lower ranges of prior estimates of the Alfv\`en surface altitude e.g. \cite{DeForest2014,Wexler2021}.  However, given the excellent validation of our source mapping for this time interval, we can go one step further and relate this empirical structure to the coronal magnetic topology.

In the top panel of figure \ref{fig:7}, we present the same field line mapping for E10 as was depicted in figure \ref{fig:5} from a 3D isometric view, and this time use the color of the Alfv\`en mach number mapped from the relevant timestamp in the PSP measurements. This shows us that the sub-alfv\`enic regions (where the colorscale goes from white to dark blue) occur over the unipolar magnetic separatrices between the individual coronal hole sources (pseudostreamers). 

\section{Discussion}\label{sec:discussion}

In this work, we presented the full cycle of efforts made to predict PSP's solar wind sources for the purpose of facilitating coordinated observations, and subsequent evaluation methods to check the accuracy. We exemplified this with the specific campaign for PSP encounter 10, which was the first orbit to reach a perihelion distance of $13.3R_\odot$. We showed that these most recent orbits provide new tools for the prediction evaluation, using the huge and rapid sampling of spatial structure, and showed that E10 was particularly accurate. We used this example in contrast to a less accurate encounter (E04) to comment on the coronal conditions when such predictions are most accurate. Lastly, we showed how the spatial structure implied by these recent direct measurements can be connected with the magnetic topology of the corona and made novel inferences about the shape of the solar alfv\`en surface as it relates to the presence of pseudostreamer topology (and resulting slow solar wind) low in the corona.

In section \ref{sec:campaigns}, we presented a detailed account of the methodology used to make footpoint predictions for each encounter of the PSP mission. Our approach is model agnostic in the sense that it regards predictions from all different models as ensemble members from which a probability distribution is fitted, and then a consensus source location is extracted as the location of maximum likelihood. 

In section \ref{subsubsec:models}, we discussed the footpoint modeling procedure and the specific models used in this work. It is worth remarking briefly on the shared limitations of these models. Although the MHD models used are much more sophisticated than the PFSS models in terms of constituent physics, they are all time-independent and driven by the same set of boundary conditions. A key assumption for the source estimates is that our model of the corona and heliosphere remains valid for at least the time it takes a solar wind plasma parcel to propagate out to PSP. While this is undoubtedly a better assumption for PSP than, for example, 1 au connectivity, it does necessarily mean there is unmodeled physics not present in our ensemble. Another missing element is that our models are all smooth and relatively low resolution, meaning they do not contain any waves or turbulence. These effects could lead to plasma parcel random-walks and therefore fieldline meandering (e.g. \citeA{Chibber2021}).

However, as we saw in section \ref{sec:evaluation}, regardless of this missing physics, our footpoint predictions under the right conditions (section \ref{subsec:e10-e04-comparison}), and especially for PSP E10, do correspond well to \textit{in situ} data at large scales. It is nevertheless worth being cognizant of these subtleties and realizing that large scale accuracy does not automatically mean accuracy at arbitrary precision (e.g. at scales much smaller than the source coronal holes). 

In section \ref{sec:e10-results} we showed an exposition of the prediction results for E10. Over nine days between inbound and outbound corotation, PSP was predicted to connect to a series of three negative polarity equatorial coronal holes, with the traversal between all three taking place in $\sim2$ days around perihelion (November 21, 2021). Prior to having any \textit{in situ} data to validate, there were already at least two reasons to be confident in the prediction. First, the ensemble of footpoints obtained from different models and boundary condition choices was tightly clustered and consistent. Second, the sources the predictions clustered on were verifiable in synchronous EUV images available at the time of the prediction.

%Discuss : 
Then, in section \ref{sec:evaluation}, we explored the process of evaluating the predictions both for E10 and more generally. In section \ref{subsec:in-situ-data} we presented \textit{in situ} timeseries data from E10 and showed how the motion of PSP distorts the data profiles when it is cast into spatial coordinates.  In section \ref{subsec:e10-eval}, we examined how the spatial coronal structure implied by the thermal plasma measurements made by PSP for E10 provided a very clear validation of the prediction in addition to predicting a consistent (negative) magnetic polarity.

In prior encounters, magnetic polarity has been our main discriminator in evaluating the predictions. Although the constraint can be somewhat stronger if there is an HCS crossing inside the encounter (where polarity in the data clearly changes from one sign to another, e.g. \citeA{Badman2022}), in general, this is a rather weak source verification method since it can only verify that the predicted source was located in the correct magnetic sector but cannot discriminate between particular coronal holes or other sources.

% This compelling model-data correspondence was shown in this work with PSP's encounter 10 (November 2021) for which the data-model comparison was especially clear and simple.  Further, the coronal holes were individually distinctive in that the first two had significant spatial extend in longitude and latitude, while the third was much smaller, thin and extended. The prediction was robust and consistent across different models, as well as variation in input magnetogram to the same model (Figure \ref{fig:3}). Additionally, the coronal holes could be independently verified to exist via direct EUV observations available at the time the prediction was made (e.g. figure \ref{fig:3}). The subsequently obtained \textit{in-situ} data when cast to spatial coordinates readily showed that 1) the magnetic polarity was continuously negative, but more powerfully 2) that PSP encountered a series of three solar wind velocity streams with marked transitions which corresponded very well to the predicted locations (and therefore timings) of the predicted connections. Additionally, 

 However, the velocity stream structure correspondence is a much more powerful constraint as it directly addresses the structure and morphology of the sources. The first two streams had moderately fast wind at their peaks, while the third was continuously very slow. The first two coronal holes accordingly were relatively large and shaped such that they clearly had an exterior boundary and a central region where the field undergoes less non-radial expansion. The third, meanwhile, was thin and extended, suggesting almost all open flux emanating from it necessarily has a high expansion factor. The relationships between these coronal holes and the peak solar wind speeds conform well with the expectations of empirical relations between expansion factors and solar wind speed \cite{Wang1990,Arge2000} as well as the distance from the coronal hole boundary \cite{Riley2001,Arge2003}.
  Further, the minima in velocity between the streams align well with unipolar separatrices, both while traversing pseudostreamers (around 15$^o$ and 100$^o$) and in approaching (but not crossing) the HCS around $60^o$, which is consistent with prior associations between such separatrices and slow solar wind (e.g. \citeA{Antiochos2011,Wang2012}), especially slow solar wind dominated by alfv\`enic fluctuations (e.g. \citeA{Panasenco2020}) which is nearly ubiquitous in PSP observations \cite{Bale2019}.
  
  This comparison method as a whole is a major novelty of PSP's most recent orbits that relies on the fact that PSP crosses a huge range of longitudinal distance around the Sun in a very short time. Earlier in the mission, PSP's perihelion hovered for 2 weeks over a very small region of the Sun \cite{Bale2019,Badman2020}. Other missions have only ever moved across the Sun's surface in slow retrograde motion (similar to the Earth) and at much larger heliocentric distances meaning temporal evolution of the corona and stream interaction and evolution are much harder to disentangle from the original coronal source structure.

In section \ref{subsec:e10-e04-comparison}, we investigated under which conditions our footpoint predictions were accurate with reference to E10 (for which we had excellent validation of the predictions), and E04 (for which the predictions were extremely uncertain). The two situations were interesting to contrast in that E04 occurred during extremely quiet solar minimum conditions, while E10 occurred later in the solar cycle. The key difference was that since there is intrinsic noise in the position of the heliospheric current sheet when considering model results from different magnetograms and model codes, a flat, near-equatorial current sheet means the footpoints of a near-equatorial spacecraft such as PSP are quite unstable: small amounts of variation can cause the HCS sepatatrix to cross the spacecraft trajectory unpredictably, and change the footpoint location drastically.  By contrast, in E10 the HCS was substantially warped and located well away from where PSP spent the encounter. This meant that small amounts of model-model variation barely changed the footpoints since the source coronal holes appeared near-universally in all prediction ensemble members. Additionally, while there was solar activity on the Sun during E10, PSP's orbit was fortuitously aligned with quiet regions without significant active region or transient activity. 

%This is similar to what was found for PSP's first sub-alfvenic crossing \cite{Kasper2021}. It is also similar to recent association with overexpanded field lines from the edge of the coronal holes of \citeA{Liu2023-preprint} in regions they term ``low-Mach number boundary layers (LMBLs). We remark that the particular sub-alfv\`enic intervals here are driven primarily by a drop in the local solar wind speed, rather than a reduction in the alfv\`en speed via magnetic or density variations, which is slightly different to the regions studied by \citeA{Liu2023-preprint} where they find density variation is also important.

Finally, in section \ref{subsec:alfven-surface}, we explored how the combination of \textit{in situ} data from PSP's unique orbit, and source mapping for E10 gave us insight into the global structure of Alfv\'en surface. Specifically, we associated the two sub-alfv\'enic intervals of the encounter with unipolar separatrices (pseudostreamers) between neighboring solar wind streams, similar to the first such sub-alfv\`enic interval explored in \citeA{Kasper2021}. It is not a surprising association since pseudostreamers are generally thought to be a contributing source of the slow solar wind \cite{Wang2012, Wang2019a, Wang2019b} and do not have a current sheet so maintain a strong B field throughout the structure. It is also a similar association to that highlighted recently by \cite{Liu2023-preprint} who associate earlier sub-alfv\`enic crossings of PSP with overexpanded field from the edge of coronal holes, a structure they term a ``low Mach-number boundary layer'' (LMBL). We also comment that the radial scans on the inbound and outbound portions of the orbit allow the same portion of the surface to be estimated over an extended period of time. Our scaling to reconstruct $R_A$ resulted in a surface which was roughly constant, however, there was still a finite thickness due to noise in the input data. This will be interesting to compare in future work with turbulent expectations of the thickness of the Alfv\`en surface \cite{Chibber2022} and allowing the solar wind to accelerate self-consistently (e.g. \citeA{Dakeyo2022}).

The two individual sub-alfv\`{e}nic intervals bear some discussion. The first one which occurs between the two fast wind streams is relatively narrow (about $5^o$ longitude) and the mach number only reaches $\sim0.9$, implying it is a marginal crossing. Additionally, it occurs at the same time as PSP is thought to have crossed through the leg of a slow-moving CME released earlier \cite{McComas2023}, thus it is likely the Alfv\`en surface location is perturbed by this transient disturbance. Since CMEs are typically faster than the ambient solar wind they plow through, it is possible the Mach number would drop further if PSP traversed the separatrix earlier.

The second interval is a more substantial sub-alfv\`enic crossing spanning over $20^o$ of longitude and with $M_A$ reaching as low as 0.1. Additionally, embedded in the stream is a sharp excursion back into the super-alfv\`enic regime driven by a sudden drop out in the magnetic field. The longitude where this occurs (and the turbulent properties therein, \citeA{Zhao2022}) is consistent with PSP having a local close approach to the HCS, depending on the model chosen from the prediction ensemble (see e.g. the difference between the grey and black HCS in the bottom panel of figure \ref{fig:6}). Therefore PSP may be sampling a relatively non-trivial region of plasma in which pseudostreamers and the solar helmet streamer interact with each other and create a complex undulation in the Alfv\`en surface.

With this Alfv\`n surface investigation, we are showing that magnetic topology low in the corona has a directly traceable imprint on the near-Sun solar wind - namely the outward protrusions in the Alfv\`{e}n surface, over a large range of solar longitude as probed by the unique orbit of PSP. This traceability between sources to make powerful inferences about the corona and heliosphere has also recently been explored by \cite{Bale2023}. Those authors use the clearly defined first two coronal holes studied here to examine the acceleration mechanism of the fast solar wind, and were able to convincingly tie the predicted solar wind signatures of interchange reconnection of mixed polarity field within the open field regions to \textit{in situ} signatures present in those two solar wind streams.

\section{Conclusions}\label{sec:conclusions}

Here we summarize our overall findings from this study.

We demonstrated our methodology for predicting the magnetic footpoints of PSP ahead of its perihelia passes, including our method to improve accuracy and quantify uncertainty via establishing a consensus footpoint from ensembles of predictions from different models and boundary conditions. We illustrated via studying PSP Encounter 10, that such an approach can and does work, since when the footpoints are most accurate, the predictions across different modeling approaches converge to the same sources.

The orbital dynamics of PSP’s most recent perihelia enable the interpretation of the \textit{in-situ} timeseries data as a spatial prograde cut through a substantial portion of the inner heliosphere and outer corona in a very short time (over 120$^o$ in three days). This rapid spatial cut will only grow in scale as the mission perihelion further shrinks.

This spatially-interpreted data enables a novel and powerful way to evaluate source mapping predictions for PSP: The thermal plasma components measured by PSP/SWEAP \cite{Kasper2016} exhibit clear longitudinal stream structure. This structure can be directly traced to the transitions between different sources. Enabling this is:

\begin{enumerate}
    \item The large spatial scale of the perihelia, meaning significantly distinct sources are sampled sequentially.
    \item The rapid motion, meaning that the coronal structure does not have time to evolve substantially  between sources, and therefore is appropriately modeled by a single time-independent model.
    \item The extremely close distance of PSP to the Sun, meaning that the solar wind is much less processed and the streamlines are all near radial so stream interactions do not disrupt the stream structure to a significant extent.
    \item The relatively favorable alignment of the orbit with the Earth, allowing the most relevant regions of the photosphere to be observed near synchronously with the magnetograms which are the fundamental boundary condition of the individual models.
\end{enumerate}

For E10 specifically, this correspondence was especially clear and took the form of three clear solar wind velocity streams corresponding to three distinct coronal hole sources of the same (negative) polarity. Pseudostreamer separatrices between the different unipolar sources corresponded to local minima in solar wind speed and additionally locations where PSP’s crossed the Alfv\`en surface. The different sizes and shapes of the coronal holes correlated well with the wind speeds (fast wind for the first two and slow wind for the third).

Next, using E10 as a case study of ``successful footpoint predictions'', the conditions under which such predictions are accurate were illustrated by contrasting with Encounter 04 (January 2020), when the prediction quality was much lower, even though there was almost perfect solar minimum conditions. In both cases, there was very low solar activity at the longitudes that PSP explored during in encounter, however, this was not a sufficient condition for accurate footpoints. An additional ingredient was the presence of a significant inclination or warping in the HCS , such that small day-to-day or magnetogram-to-magnetogram model variation does not significantly move magnetic separatrices with respect to PSP's trajectory.

For the E04 results where the footpoints were strongly bifurcated, it was observed that footpoints were more similar among model results driven by the same magnetogram than by different models driven by the same magnetogram, implying the input boundary condition is at least as important as the physics of the model for the application of field line tracing (see also \citeA{Riley2006,Badman2022}).

Lastly, moving beyond their role in source validation, the \textit{in situ} spatial cuts themselves constitute large-scale slices through the global structure of the outer corona and inner heliosphere. Combined with the footpoint mapping, we can associate such structure with underlying magnetic topology. We demonstrated this with the measurements of the Alfv\'en Mach number for E10 and found that it implies an average surface just below the current closest approach (13.3$R_\odot$), but also a wrinkled surface with narrow protrusions which cross PSP’s trajectory (see also \citeA{Verscharen2021}, \citeA{Liu2023-preprint}). By associating these protrusions with the well-validated source mapping, we observed they appear to occur over slow solar wind streams emanating from the tips of pseudostreamers or at least the overexpanded edges of coronal holes, as was reported for the first such crossing \cite{Kasper2021}.

\section{Future Work}\label{sec:future-work}

PSP's prime mission includes at least ten further orbits (at the time of writing), the latter eight of which will all be even closer and therefore consist of an even larger spatial cut through coronal structure. Therefore we anticipate repeating the footpoint prediction and validation exercise presented here for all such orbits.  There will be further iterations to make these as accurate, useful and practical as possible.

For example, as shown by the pipeline in the associated GitHub repository for this work \cite{BadmanZenodo2023}, the analysis required to produce the predictions is very close to being automated. Therefore we aim to complete this process and make available the pipeline as a python script. Some human interaction is still needed to produce text descriptions of the predictions to account for such factors as solar activity or multiple source centroids (\ref{appendix:sec:practical}). 

Additionally, an as yet untapped direction for improving the accuracy of our footpoint predictions is to use complementary near real-time observations from 1 au spacecraft and Solar Orbiter (SolO; \citeA{Muller2020}) where available to narrow down the parameter space of the models used in the footpoint ensembles. Specifically, magnetic polarity measurements of the most recent orbits to cross the Carrington longitudes of PSP's upcoming orbits can be used to constrain the position of the heliospheric current sheet and reject any model results which are clearly inconsistent with this. As another example, for cases (such as E10) where the predicted sources are coronal holes with specific morphology, the most recent direct EUV observations can be used to confirm or refute their existence, or to choose ensemble members where the open field morphology is most similar to the EUV observations and weight them higher.

A further key issue for producing useful coordinated observations which was not deeply addressed in the present work is the issue of transit timing from Sun to PSP. One objective of such observations is to directly observe the injection of impulsive signatures on open field lines at the coronal base and associate them directly with \textit{in situ} signatures at PSP. The time lag between such emission and receipt depends on the disturbance (whether it is advected with the solar wind, or propagates at an MHD wave speed in the solar wind), but can be of order hours to tens of hours. At peak angular velocity, PSP can move from one source to another \textit{at this same timescale}. Thus, when providing a footpoint and associated timestamp, estimating the transit time can be important. Since E10, this consideration has informed an additional component in the prediction files of an ``emission time'' and ``receipt time'' with an unsophisticated ballistic estimate. For further encounters, especially those closer to the Sun, this issue will become even more pronounced, so the estimate should be improved using, for example, the most recent available SolO data, to constrain a Parker solar wind curve and integrate to provide an accurate transit time. 

As mentioned in section \ref{sec:discussion}, time-dependence of the models themselves is also currently not accounted for other than in the updating of the magnetic map boundary conditions as often as possible. Time-dependent phenomena such as waves and turbulence or just the development of stream interactions is an obvious direction to improve the modeling input itself to these procedures. Especially valuable would be using codes that directly follow test particles or plasma parcels through the model, rather than relying on smooth magnetic fields. While such modeling is vastly more computationally expensive than the tracing reported in this work, even one or two examples which can be compared to the results of time-static field line tracing would allow better estimates of its accuracy and precision in the absence of time-dependent effects.

Lastly, there are numerous interesting directions to pursue with regard to empirically constraining the alfv\`en surface, a key boundary mediating between coronal and solar wind physics (e.g. \citeA{Weber1967}). Specifically, the scaling we used ($R_A = R_{PSP}/M_A$) following \citeA{Liu2021,Liu2023-preprint} should be revisited and performed more carefully with the most recently available information about how the solar wind quantities scale with radius e.g. \cite{Dakeyo2022,Halekas2022}. This scaling yielded outward protrusions which are likely unphysically high, since they imply PSP would have measured sub-alfv\`enic excursions as early at its first perihelion, which it did not. There are now several orbits for which curves such as that depicted in figure \ref{fig:7} can be derived for and the source regions compared. Further, intervals such as the radial scans, or conjunctions with spacecraft can be used to investigate how stable $R_A$ is in time and how universal the radial scaling is. Lastly, as predicted here and in \citeA{Liu2021,Liu2023-preprint}, the latter orbits of the PSP mission are likely to reach the average $R_A$ value implied by our scalings, it may imminently be possible to contrast differences crossings of protrusions over special solar wind streams with samples from inside the `normal sub-alfv\`enic' corona.

\appendix
\section{Practical Aspects of the Prediction Campaigns}\label{appendix:sec:practical}

In this appendix, we discuss several practical considerations salient to the prediction campaigns and in how we disseminate the prediction information to make it useful for remote observers.
 
\subsection{Campaign Scope}\label{appendix:subsec:scope}

The consensus fitting procedure described in section \ref{subsec:consensus} is iterated through all the time bins in the provided prediction datasets to provide a timeseries of source locations proceeding into the future. In a typical campaign, model results are produced on a given day using the most recent magnetogram boundary conditions available. It is therefore anticipated that the longer the time horizon of the predictions, the less accurate they will be due to future time-evolution of the boundary conditions to the model. For this reason, several updates to the predictions are issued during a typical encounter.

Additionally, since the primary goal of the predictions to produce observational targets for Earth-based observers, we are mainly concerned with times for which the predicted footpoints are ``on-disk''. Different PSP encounters have very different on-disk phases, with some encounters occurring mostly behind the limb, some occurring fully in view of the earth, and others (most commonly) either going behind the West limb or appearing from behind the East limb part way through the encounter. In the most recent orbits, this latter situation is almost always the case by virtue of PSP traversing more than a third of the way around the Sun during the encounter. 

These considerations affect the scope of the footpoint campaign and the frequency of updates. In situations where footpoints are mostly on disk, updates are given more frequently and over a longer time interval, whereas in the converse situation, only a few updates are given and over a shorter time range.  The predicted footpoints are usually truncated at the point where the footpoints go behind the Earth-facing limb, or once PSP is much further from the Sun than perihelion.

The on-disk phases are determined ahead of time via ballistic projection estimates \cite{Badman2020} using the PSP spice kernels. Example figures and animations showing these determinations are hosted at the following url: \url{https://sppgway.jhuapl.edu/encounters}.

\subsection{Prediction Dissemination}\label{appendix:subsec:dissemination}

In order to communicate the prediction results in a timely fashion to observers, as part of our prediction pipeline, we have defined a typical set of files produced in each update. This includes a CSV file encoding of the predicted source locations (latitude and longitude, as well as solar disk coordinates $\theta_x,\theta_y$), figures which depict the source location and uncertainty ellipse both in a Carrington/synoptic map projection and a solar disk projection, and lastly a manually written text narrative where the consensus motion on the disk is described, and any non-trivialities are discussed.

In more recent encounters (since E12 onwards), the CSV files have included two timestamps to account for the two intervals of primary interest to making PSP-solar connections: time of arrival at PSP of plasma parcels from a given source at PSP, and the new addition of the time of emission of the source \textit{at the footpoint} which we estimate by assuming some constant ambient solar wind speed, and constant propagation speed all the way back to the Sun. Additionally, since the solar disk rotates between these times, although the Carrington longitudes connecting both these events (emission at the source, and arrival at PSP) are fixed in time (within the assumptions of time-independent modeling), the on-disk position of the source is slightly different. Therefore a second value of $(\theta_x,\theta_y)$ is provided for this earlier timestamp as well.

The emission time is subject to significant uncertainties since the true solar wind speed at PSP is unknown at the time of prediction, and moreover the solar wind accelerates so a ballistic estimate is necessarily and underestimate of the propagation time, and better constraints on it will be the subject of future work in these campaigns. 

The platform over which these predictions are communicated also varies with the campaign scope. For partially on-disk orbits, with fewer updates, the predictions are conveyed over email to interested observers via a mailing list maintained by the PSP management team, while for the more involved fully on-disk occasions, campaigns are managed via the Whole Heliosphere and Planetary Interactions group who host and archive the predictions at their website: \url{https://whpi.hao.ucar.edu/}.

\subsection{Fitting Failures}\label{appendix:subsec:fitfail}

As an operational procedure, we typically produce predictions in a relatively time-constrained manner. As such, when running the fitting procedure describe in section \ref{subsec:consensus} we need to be able to handle occasions when it fails. The most common failure mode we have observed is when the ensemble is bi-modal or multi-modal, which typically occurs in the vicinity of predicted HCS crossings. Another apparent failure occurs when the distribution only consists of a few points or is too clustered, in which case the fitting software we use \cite{Fraenkel2014} has numerical difficulties, and a fit does not converge. 

In the first case, we typically find the different peaks in the ensemble are located at very different latitudes. As a solution, we therefore take the most probable peak (the peak which contains the most ensemble members) and exclude the other from the fit via masking by latitude. When reporting the predictions, although our consensus data sets show this primary peak, we usually mention the coronal feature corresponding to the other peaks in the text narrative.

In the second case where the ensemble is too tightly clustered, the purpose of the Kent distribution in capturing the uncertainty is less necessary since we can simply regard the uncertainty as ``small''. Therefore when this fitting failure mode occurs we do not attempt to fit again but instead, simply take the median longitude and latitude from the ensemble and use this as the consensus location for that time window.

%%%%%%%%%%%%%%%%%%%%%%%%%%%%%%%%%%%%%%%%%%%%%%%%%%%%%%%%%%%%%%%%
%
% Optional Glossary, Notation or Acronym section goes here:
%
%%%%%%%%%%%%%%
% Glossary is only allowed in Reviews of Geophysics
%  \begin{glossary}
%  \term{Term}
%   Term Definition here
%  \term{Term}
%   Term Definition here
%  \term{Term}
%   Term Definition here
%  \end{glossary}

%
%%%%%%%%%%%%%%
% Acronyms
%   \begin{acronyms}
%   \acro{Acronym}
%   Definition here
%   \acro{EMOS}
%   Ensemble model output statistics
%   \acro{ECMWF}
%   Centre for Medium-Range Weather Forecasts
%   \end{acronyms}

%
%%%%%%%%%%%%%%
% Notation
%   \begin{notation}
%   \notation{$a+b$} Notation Definition here
%   \notation{$e=mc^2$}
%   Equation in German-born physicist Albert Einstein's theory of special
%  relativity that showed that the increased relativistic mass ($m$) of a
%  body comes from the energy of motion of the body—that is, its kinetic
%  energy ($E$)—divided by the speed of light squared ($c^2$).
%   \end{notation}

\section{Open Research}

The predicted footpoint ensembles used in figures \ref{fig:2}, \ref{fig:3}, \ref{fig:5} and \ref{fig:6} are hosted in a GitHub repository accompanying this paper \cite{BadmanZenodo2023}, which also contains a python jupyter notebook which reproduces all the figures shown in this paper and includes the analysis steps of producing the consensus footpoints. 

\subsection{Datasets}

All data other than the footpoint ensembles are publicly available from the relevant NASA, mission or instrument webpages, and their access is performed inline by the aforementioned jupyter notebook.

The Parker Solar Probe FIELDS \cite{Bale2016,Pulupa2017} and SWEAP data \cite{Kasper2016,Livi2022} used for in the study to constrain and validate the footpoint predictions are archived at the Space Physics Data Facility via \url{https://spdf.gsfc.nasa.gov/pub/data/psp/} and are publically available \cite{Candey2021}. These data are accessed programmatically in this work in python by \verb+pyspedas+ (see below).

The Solar Dynamics Observatory Atmospheric Imager Assembly data \cite{Lemen2012} used to contextualize footpoint measurements in this work are archived by the Virtual Solar Observatory \url{https://nso.edu/data/vso/}, and the Joint Science Operations Center \url{http://jsoc.stanford.edu/} and are publicly available. These data are accessed programmatically in this work in python via \verb+solarsynoptic+ (see below).

The ADAPT magnetic field maps \cite{Arge2010,Arge2011,Arge2013,Hickmann2015} used to run coronal models in this work are archived by the National Solar Observatory at \url{https://gong.nso.edu/adapt/} and are publicly available. Data is accessed directly in this work via http requests at the above URL.

\subsection{Software}

All software used in this paper are either included as helper functions in the linked repository \cite{BadmanZenodo2023} or use open source python libraries:

Version 5.1 of the \verb+astropy+ is used throughout this work for coordinate and frame transformations and working with units. It is preserved at doi:10.5281/zenodo.6579729, available via open access and developed openly at \url{https://github.com/astropy/astropy}. \cite{astropy:2013,astropy:2018,astropy:2022}

Version 0.2.0 of the \verb+astrospice+ is used throughout this work for spacecraft trajectory calculation. The persistent link is \url{https://pypi.org/project/astrospice/0.2.0/}. It is developed openly at \url{https://github.com/astrospice/astrospice}. 

Version 4.3.5 of \verb+func_timeout+ is used for error handling in the consensus fitting procedure. The persistent link is \url{https://pypi.org/project/func-timeout/4.3.5/}. It is developed openly at \url{https://github.com/kata198/func_timeout}. 

\verb+kent_distribution+ is used for establishing consensus fits in this work. It is openly developed at \url{https://github.com/edfraenkel/kent_distribution}. \cite{Fraenkel2014}. An edited version of this repository as used here to allow it to funciton in \verb+python3+ is available via \citeA{BadmanZenodo2023}.

Version 3.5.2 of \verb+matplotlib+ is used for all figures presented in this work. It is preserved at doi:10.5281/zenodo.6513224, available via open access and developed openly at \url{https://github.com/matplotlib/matplotlib}. \cite{Hunter:2007}

Version 1.1.2 of \verb+pfsspy+ is used for PFSS coronal modeling and field line tracing throughout this work. It is preserved at doi:10.5281/zenodo.7025396, available via open access and developed openly at \url{https://github.com/dstansby/pfsspy}. \cite{Stansby2020}

Version 1.4.18 of \verb+pyspedas+ is used for PSP data access. The persistent link is  \url{https://pypi.org/project/pyspedas/1.4.18/}. The project is developed openly at \url{https://github.com/spedas/pyspedas}.  \cite{Angelopoulos2019}

Version 1.9.1 of \verb+scipy+ is used for data interpolation in this work. It is preserved at doi:10.5281/zenodo.7026742, available via open access and developed openly at \url{https://github.com/scipy/scipy}. \cite{Virtanen2020}

Version 5.1.2 of \verb+spiceypy+ is the engine behind \verb+astrospice+.  It is preserved at doi:10.5281/zenodo.7204403 and is available via open access and developed openly at \url{https://github.com/AndrewAnnex/SpiceyPy}. \cite{Annex2020}

Version 4.1.0 \verb+sunpy+ is used for enumerable tasks in this manuscript, most notably frame transformations, plotting and doing world to pixel coordinate conversions. It is preserved at doi:10.5281/zenodo.7314636 and is available via open access and developed openly at \url{https://github.com/sunpy/sunpy}. \cite{sunpy_community2020}

% AGU requires an Availability Statement for the underlying data needed to understand, evaluate, and build upon the reported research at the time of peer review and publication.

% Authors should include an Availability Statement for the software that has a significant impact on the research. Details and templates are in the Availability Statement section of the Data and Software for Authors Guidance: \url{https://www.agu.org/Publish-with-AGU/Publish/Author-Resources/Data-and-Software-for-Authors#availability}

% It is important to cite individual datasets in this section and, and they must be included in your bibliography. Please use the type field in your bibtex file to specify the type of data cited. Some options include Dataset, Software, Collection, ComputationalNotebook. Ex: 
% \\
% \begin{verbatim}

% @misc{https://doi.org/10.7283/633e-1497,
%   doi = {10.7283/633E-1497},
%   url = {https://www.unavco.org/data/doi/10.7283/633E-1497},
%   author = {de Zeeuw-van Dalfsen, Elske and Sleeman, Reinoud},
%   title = {KNMI Dutch Antilles GPS Network - SAB1-St_Johns_Saba_NA P.S.},
%   publisher = {UNAVCO, Inc.},
%   year = {2019},
%   type = {dataset}
% }

% \end{verbatim}

% For physical samples, use the IGSN persistent identifier, see the International Geo Sample Numbers section:
% \url{https://www.agu.org/Publish-with-AGU/Publish/Author-Resources/Data-and-Software-for-Authors#IGSN}
%%%%%%%%%%%%%%%%%%%%%%%%%%%%%%%%%%%%%%%%%%%%%%%

\acknowledgments

 Parker Solar Probe was designed, built, and is now operated by the Johns Hopkins Applied Physics Laboratory as part of NASA's Living with a Star (LWS) program (contract NNN06AA01C). Support from the LWS management and technical team has played a critical role in the success of the Parker Solar Probe mission. The FIELDS and SWEAP experiments on Parker Solar Probe spacecraft were designed and developed under NASA contract NNN06AA01C. 
 
The ADAPT model development is supported by Air Force Research Laboratory (AFRL), along with AFOSR (Air Force Office of Scientific Research) tasks 18RVCOR126 and 22RVCOR012. This work utilizes data produced collaboratively between AFRL and the National Solar Observatory (NSO). The views expressed are those of the authors and do not reflect the official guidance or position of the United States Government, the Department of Defense (DoD) or of the United States Air Force. The appearance of external hyperlinks does not constitute endorsement by the DoD of the linked websites, or the information, products, or services contained therein. The DoD does not exercise any editorial, security, or other control over the information you may find at these locations.

 The authors thank the Whole Heliosphere and Planetary Interactions group who coordinated the first iteration of the now-regular footpoint predictions presented in this work and have regularly supported the dissemination and hosting of the results to remote observers. 

 RCA acknowledges support from NASA grant 80NSSC22K0993.

 PR gratefully acknowledges support from NASA (80NSSC18K0100, NNX16AG86G, 80NSSC18K1129, 80NSSC18K0101, 80NSSC20K1285, 80NSSC18K1201, and NNN06AA01C), NOAA (NA18NWS4680081), and the U.S. Air Force (FA9550-15-C-0001).

 STB thanks Carlos Braga for useful discussions on the CME leg crossing studied in \citeA{McComas2023} which coincides with the first sub-alfv\`enic interval in this work.

 JLV acknowledges support from NASA PSP-GI 80NSSC23K0208.

TKK and NVP acknowledge support from NASA grant 80NSSC20K1453, NSF/NASA SWQU grant 2028154, and the PSP mission through the UAH–SAO agreement SV4-84017. TKK acknowledges support from AFOSR grant FA9550-19-1-0027.
% This section is optional. Include any Acknowledgments here.
% The acknowledgments should list:\\
% All funding sources related to this work from all authors\\
% Any real or perceived financial conflicts of interests for any author\\
% Other affiliations for any author that may be perceived as having a conflict of interest with respect to the results of this paper.\\
% It is also the appropriate place to thank colleagues and other contributors. AGU does not normally allow dedications.

%% ------------------------------------------------------------------------ %%
%% References and Citations

%%%%%%%%%%%%%%%%%%%%%%%%%%%%%%%%%%%%%%%%%%%%%%%
%
% \bibliography{<name of your .bib file>} don't specify the file extension
%
% don't specify bibliographystyle

% In the References section, cite the data/software described in the Availability Statement (this includes primary and processed data used for your research). For details on data/software citation as well as examples, see the Data & Software Citation section of the Data & Software for Authors guidance
% https://www.agu.org/Publish-with-AGU/Publish/Author-Resources/Data-and-Software-for-Authors#citation

%%%%%%%%%%%%%%%%%%%%%%%%%%%%%%%%%%%%%%%%%%%%%%%

\bibliography{agusample.bib}

%Reference citation instructions and examples:
%
% Please use ONLY \cite and \citeA for reference citations.
% \cite for parenthetical references
% ...as shown in recent studies (Simpson et al., 2019)
% \citeA for in-text citations
% ...Simpson et al. (2019) have shown...
%
%
%...as shown by \citeA{jskilby}.
%...as shown by \citeA{lewin76}, \citeA{carson86}, \citeA{bartoldy02}, and \citeA{rinaldi03}.
%...has been shown \cite{jskilbye}.
%...has been shown \cite{lewin76,carson86,bartoldy02,rinaldi03}.
%... \cite <i.e.>[]{lewin76,carson86,bartoldy02,rinaldi03}.
%...has been shown by \cite <e.g.,>[and others]{lewin76}.
%
% apacite uses < > for prenotes and [ ] for postnotes
% DO NOT use other cite commands (e.g., \citet, \citep, \citeyear, \citealp, etc.).
% \nocite is okay to use to add references from your Supporting Information
%

\end{document}